%% file: rl_rq_bals.tex
\documentclass[useAMS,usenatbib]{mn2e}
\usepackage{mathtools}
\usepackage{pdflscape}
\usepackage{longtable}
\usepackage{graphicx}
\usepackage{natbib}
\usepackage{lscape}
\usepackage{longtable}
\usepackage{subfig}
\usepackage{amssymb,amsmath}
\usepackage[T1]{fontenc}
\usepackage{hyperref}
\usepackage{amsmath}
\usepackage{breakurl}
\usepackage{times}

\input{commands_mnras.tex}

\title[UV Comparison of RL and RQ BAL QSOs]{Radio-Loud and Radio-Quiet BAL Quasars: A Detailed Ultraviolet Comparison}
\author[T. B. Rochais (et al.)]{T. B. Rochais$^1$, M. A. DiPompeo$^1$, A. D. Myers$^1$, M. S. Brotherton$^1$, J.C. Runnoe$^2$ \newauthor S.W. Hall$^1$ \\
$^1$ University of Wyoming, Dept. of Physics and Astronomy 3905, 1000 E. University, Laramie, WY 82071, USA \\
$^2$ Pennsylvania State University, Dept. of Astronomy and Astrophysics \& Center for Gravitational Wave Physics, 525 Davey Lab, University Park, PA 16802, USA
}

\begin{document}
\date{Accepted ?; Received ?; in original form 2014 July}

\pagerange{\pageref{firstpage}--\pageref{lastpage}} \pubyear{2014}

\maketitle

\label{firstpage}

\begin{abstract}
Studies of radio-loud (RL) broad absorption line (BAL) quasars indicate that popular orientation-based BAL models fail to account for all observations.  Are these results extendable to radio-quiet (RQ) BAL quasars?  Comparisons of RL and RQ BAL quasars show that many of their properties are quite similar.  Here we extend these analyses to the rest-frame ultraviolet (UV) spectral properties, using a sample of 73 RL and 473 RQ BAL quasars selected from the Sloan Digital Sky Survey (SDSS). Each RQ quasar is individually matched to a RL quasar in both redshift (over the range $1.5 < z < 3.5$) and continuum luminosity. We compare several continuum, emission line, and absorption line properties, as well as physical properties derived from these measurements.  Most properties in the samples are statistically identical, though we find slight differences in the velocity structure of the BALs that cause apparent differences in \CIV\ emission line properties. Differences in the velocities may indicate an interaction between the radio jets and the absorbing material.  We also find that UV \FeII\ emission is marginally stronger in RL BAL quasars.  All of these differences are subtle, so in general we conclude that RL and RQ BAL QSOs are not fundamentally different objects, except in their radio properties. They are therefore likely to be driven by similar physical phenomena, suggesting that results from samples of RL BAL quasars can be extended to their RQ counterparts.
\end{abstract}


\section{INTRODUCTION}

It is well accepted that a supermassive black hole lies at the center of every massive galaxy. In some cases, this black hole is surrounded by an accretion disk that emits powerful radiation visible across the observable universe --- these are the most luminous active galactic nuclei (AGN), the so-called quasars. However, this simple definition of a quasar does not account for the fact that many different subclasses of quasars are observed.  For example, broad absorption line (BAL) quasars represent around 20\% of the optically-selected quasar population \citep{Knigge08}. These quasars show strong blueward-displaced absorption lines in their rest-frame ultraviolet spectra \citep[e.g.,][]{Weymann91}. This indicates the presence of massive, high-velocity (several percent of the speed of light) outflows that potentially have significant effects on their environment, the host galaxy, and the evolution of the quasar itself \citep{Scannapieco04, Vernaleo06, Hopkins06}.  BAL quasars may regulate star formation rates in their host galaxies \citep{HopkinsElvis10, Leighly14} and even play a role in the formation of large-scale structure and galaxies \citep{Vernaleo06}. While understanding the nature of BAL quasars is quite important for understanding quasars and their role in cosmic evolution, so far their true nature remains an enigma. We still do not know precisely why some quasars show BALs while others do not. 

 BAL quasars themselves are further split into a variety of subclasses. The majority fall into the category of high-ionization BAL quasars (HiBALs), where the BALs are due to elements with high ionization states, such as \CIV\  $\lambda 1549$, \SiIV\  $\lambda 1397$, and \NV\  $\lambda 1240$. A smaller fraction, on the order of 1\% (of all optically selected quasars), are low-ionization BAL quasars (LoBALs), where the BALs are formed from elements with lower ionizations, such as \MgII\ $\lambda 2799$. Finally, even more rare are Iron LoBALs (FeLoBALs) that also show absorption from \FeII\  $\lambda 2380$, $\lambda 2600$, and $\lambda 2750$. 

Two main models are usually proposed to explain BAL quasars and their various subclasses. The first is based on orientation and the second on evolution. In the orientation model, BAL winds are radiatively driven into a generally equatorial outflow, away from the accretion disk symmetry axis \citep{Elvis00}.  Note that the term ``equatorial'' here is used loosely because of the complete obscuration from dust for many quasars in the true equatorial direction \citep[i.e.\ the dusty torus; e.g.,][]{Antonucci93, UrryPadovani95}. In this scenario all quasars are roughly identical; the observed differences only result from different viewing angles through non-spherically symmetric structures --- BALs are simply seen when our line of sight intersects the outflow at a relatively large angle from the accretion disc symmetry axis. This provides a simple solution for the unification of quasars, and seems to work well at low redshift and luminosity \citep[e.g. Seyfert galaxies;][]{Antonucci93}. However, while it is somewhat supported as an explanation for BAL quasars by strong similarities in the emission-line properties of radio-quiet (RQ) BAL and non-BAL quasars \citep{Weymann91}, there are also a few differences.  This includes stronger optical \FeII\  emission and weaker \OIII\  in BAL quasars, which separates BAL and non-BAL quasars along ``eigenvector 1'' \citep[EV1][]{BorosonGreen92}.  

As an alternative explanation, it has been suggested that quasars can evolve from one type to another, thus accounting for some of the observed differences between BALs and non-BALs \citep[e.g.][]{Gregg06, FilizAk12, FilizAk13}.  According to this model BALs are seen along all lines of sight until the ``blow out'' phase, during which all the gas is ejected and the quasar becomes a non-BAL quasar. This model has some support from the fact that BAL quasars appear to have redder spectra and thus more dust \citep[e.g.][]{Sprayberry92, Brotherton01, DiPompeo12}.  However, this difference does not extend to the near-infrared \citep[][]{Gallagher07}, far-infrared \citep[][]{Cao12}, or sub-mm \citep[][]{Willott03}, which is troubling for this paradigm, although \citet{DiPompeo13} did find a near-IR excess in a sample of BAL quasars.

The orientation-only model may be tested from observations in the radio, where the steepness of the radio spectrum depends on whether radio cores or lobes dominate the emission, and thus whether the jet axis is parallel or perpendicular to the line of sight \citep[e.g.][]{Wills95}. Some work has been done to show that BAL and non-BAL quasars show no significant difference between radio spectral index distributions, thus indicating that they have a similar range of viewing angles \citep{Becker00, MontenegroMontes08} and providing strong evidence against the orientation-only model. In a larger study of well-matched BALs and non-BALs selected specifically for this test, a marginally significant difference was found showing an overabundance of steep radio spectra for BAL sources \citep{DiPompeo11b}. However, the overall range of values still covered all possible viewing angles \citep{DiPompeo12}. This suggests that orientation plays a role, but alone it cannot explain all observed properties. Combined with other recent studies \citep[e.g.][]{Gallagher07, Shankar08, Allen11, Bruni12, DiPompeo13}, it seems a combination of orientation, evolution, and possibly some yet-unknown factors are at work in the BAL class.

Studies of the radio spectral index are limited to the relatively rare subclass of radio-loud (RL) BAL quasars.  Quasars are formally RL when either their ratio of radio to optical flux is large \citep[$>10$;][]{Stocke92} or they have radio luminosities greater than $10^{25}$ W Hz$^{-1}$ sr$^{-1}$ \citep{MillerPeacockMead90}. RL quasars only represent a small fraction of quasars, about 10\%.  This fraction is even lower in BAL quasars \citep{Stocke92, Brotherton98, Becker01}. As a result, it is much easier to systematically study RQ BAL quasars.  The reason that some quasars are RL and some are RQ, as well as whether or not RL and RQ quasars form fundamentally different populations is still currently under investigation \citep[e.g.][]{MillerPeacockMead90, Goldschmidt99, Jiang07, Singal13}.  

In this work, we add to the growing body of evidence that indicates that RL and RQ BAL quasars are far more similar than they are different, and that results regarding orientation from RL BAL quasar samples are extendable to RQ BAL quasars.  For example, both \citet{Brotherton05} and \citet{Kunert09} found that the X-ray properties of RL and RQ BAL quasars are quite similar, and \citet{Welling14} found that the variability properties of RL and RQ BAL quasars are generally indistinguishable.  \citet{Runnoe13} found that RL BAL quasars are very similar to RQ BAL quasars in their optical emission-line properties, and RL BAL quasars are more ``BAL-like'' than ``RL-like'' in their EV1 properties.  \citet{Bruni14} find that the rest-frame optical properties of RL and RQ BAL quasars, and physical parameters derived from them, are identical.  Here, we examine the rest-frame ultraviolet (UV) spectral properties of a well-matched sample of 73 RL and 473 RQ BAL quasars, and search for significant differences in their properties.  We focus on emission line, continuum, absorption line, and derived physical properties.

\begin{table}
  \centering
  \caption{RQ BAL quasars selected for the comparisons.}
  \label{table:info}
  \begin{tabular}{lcccc}
    \hline
Object Name (SDSS J) & RA &  DEC  &  $z$ & Type \\
\hline
155338.20+551401.9&  238.4092&   55.2339&      1.64&HiBAL\\
122829.98+520241.8&  187.1250&   52.0450&      3.03&HiBAL\\
095422.68+524903.8&  148.5945&   52.8177&      2.34&HiBAL\\
104612.99+584719.0&  161.5542&   58.7886&      3.04&HiBAL\\
121644.20+600845.3&  184.1842&   60.1459&      1.85&HiBAL\\
074213.49+211341.1&  115.5562&   21.2281&      1.83&HiBAL\\
092015.68+350040.5&  140.0654&   35.0113&      1.92&HiBAL\\
093846.77+380549.8&  144.6949&   38.0972&      1.83&HiBAL\\
145250.76+434555.5&  223.2115&   43.7654&      1.72&LoBAL\\
142437.67+394535.5&  216.1570&   39.7599&      2.18&HiBAL\\
\hline
  \end{tabular}
  
\raggedright{
    A sample of the table is provided here for illustration of its content --- the complete table can be found in the online version.
   }
\end{table}

\section{SAMPLE}

\subsection{Radio-Loud BAL Quasars}
The RL sample is taken from \citet{DiPompeo11b}, and we refer the reader there for complete details.  We provide a brief summary here.  We begin with the BAL catalog of \citet[][hereafter G09]{GibsonCatalog09}, which is drawn from SDSS data release 5 (DR5) \citep{SDSSDR5}, keeping objects with $z > 1.5$ in order to include \CIV\ in the SDSS spectral window.  This is necessary in order to unambiguously identify BAL quasars.  Matches are searched for in the Faint Images of the Radio Sky at Twenty-Centimeters survey  \citep[FIRST;][]{BeckerFIRST95}, keeping matched sources with integrated flux densities of 10 mJy. \citet{DiPompeo12} analyzed the UV spectral properties of this sample, and discarded one object with a line-locked \CIV\  doublet that was mistakenly included as a BAL.  The final sample contains 73 bona-fide RL BAL quasars over the range $1.5 < z < 3.5$, 11 of which are Lo/FeLoBALs ($\sim$15\%).

\subsection{Radio-Quiet BAL Quasars}
The RQ BAL quasar sample also uses the G09 catalog, starting with objects that have no counterpart (within 10\arcsec) in the FIRST catalog.  We then search this subset for objects that are well-matched to individual RL objects using an iterative procedure.  For each RL object, we search for RQ sources matched within 1\% of galactic extinction-corrected $i$-band magnitude and redshift (these parameters combined give well-matched luminosities).  This percentage is increased in steps of 0.5\%, until at least four matches are found for each RL object --- the maximum percentage is 5\%.   A lower-limit is imposed on the redshifts ($z > 1.5$) to ensure that \CIV\  remains in the spectral window, as with the RL sample. 

Despite their inclusion in the G09 catalog we visually inspect all of these spectra to ensure we agree with their BAL classification, as borderline cases can be quite ambiguous, and to make our own classifications as HiBAL, LoBAL, or FeLoBAL.  Three people performed the inspection, and we discard a small number of objects that are questionable.  In the end, we are left with 473 RQ BAL quasars, 53 of which are Lo/FeLoBAL quasars ($\sim$12\%).  We point out that using the classifications from for example the catalog of \citet{ShenCatalog11}, only four objects are flagged as LoBAL quasars, highlighting the need for caution when using BAL flags from large, automated catalogs.  The full list of RQ objects and their general properties is presented in Table 1.  Figure 1 illustrates that the redshifts and luminosities of the RL and RQ samples are well-matched.

One potential concern in our sample selection is that FIRST is flux limited to above 1 mJy, and at higher redshifts some of the apparently RQ sources could be intrinsically RL but below the detection limit of the survey.  To test how significant this contamination may be, we calculate upper limits to both the radio-loudness parameter \citep[$R^*$;][]{Stocke92} and the radio luminosity at 5 GHz for each object in the RQ sample, assuming a radio spectral index $\alpha_r=-0.5$.  We use a cosmology where $\Omega_M = 0.27$, $\Omega_{\Lambda} = 0.73$, and $H_0 = 71$ km s$^{-1}$ Mpc$^{-1}$ \citep{Komatsu11} for the luminosity calculations.  Values of $f_{2500}$ from the \citet{ShenCatalog11} catalog are adopted in the calculation of $R^*$.  Distributions of these upper limits for the RQ sample are shown in Figure~\ref{fig:limits}.  Only 11 sources have upper limits of $\log R^* > 1$, and none have $L_{5\textrm{GHz}} > 10^{25}$ W Hz$^{-1}$ sr$^{-1}$. We conclude that there will be minimal effects on our results from intrinsically radio-loud objects that are not detected in FIRST.

\begin{figure}
\centering
\hspace{0cm}
   \includegraphics[width=5.5cm]{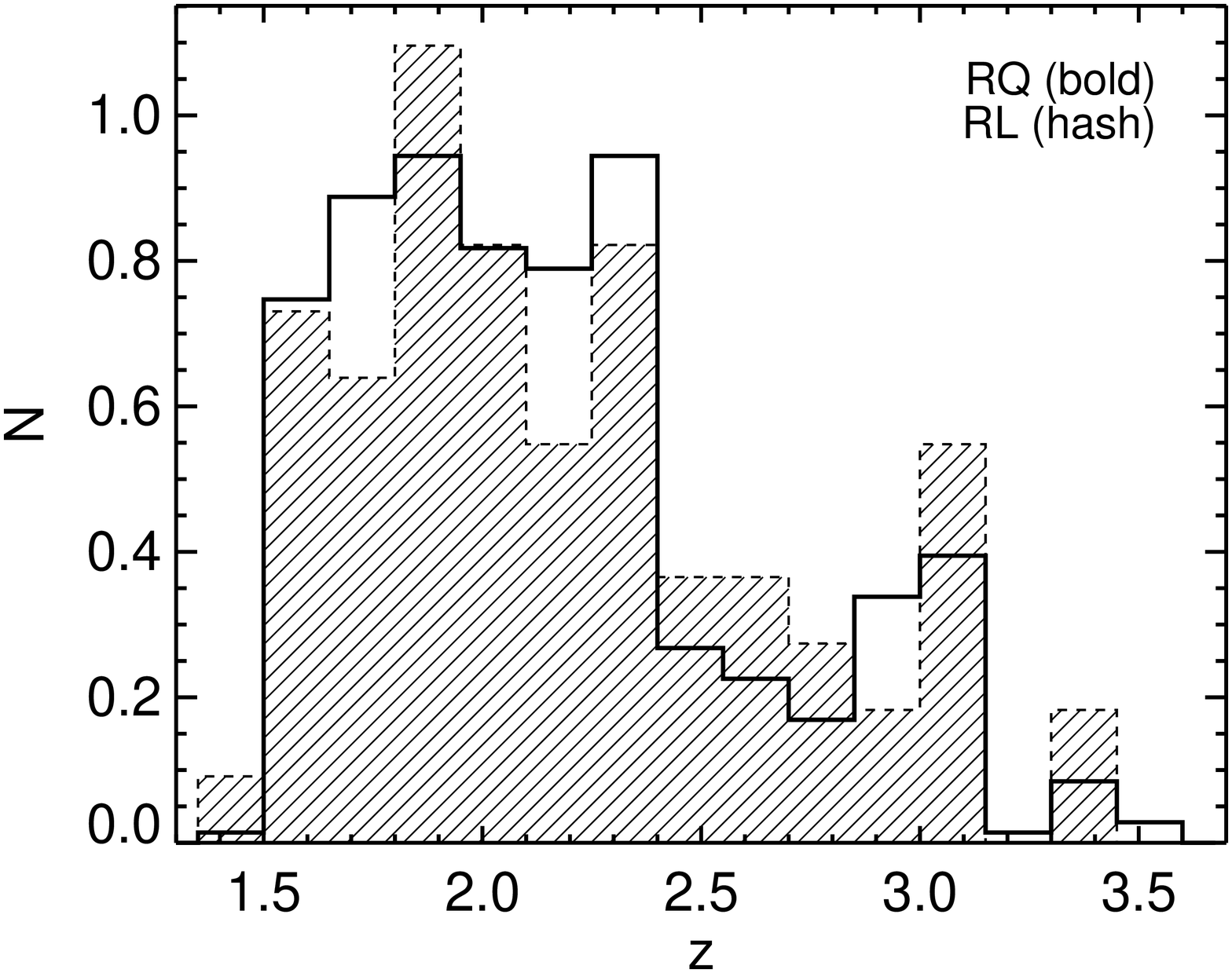}
   \includegraphics[width=5.5cm]{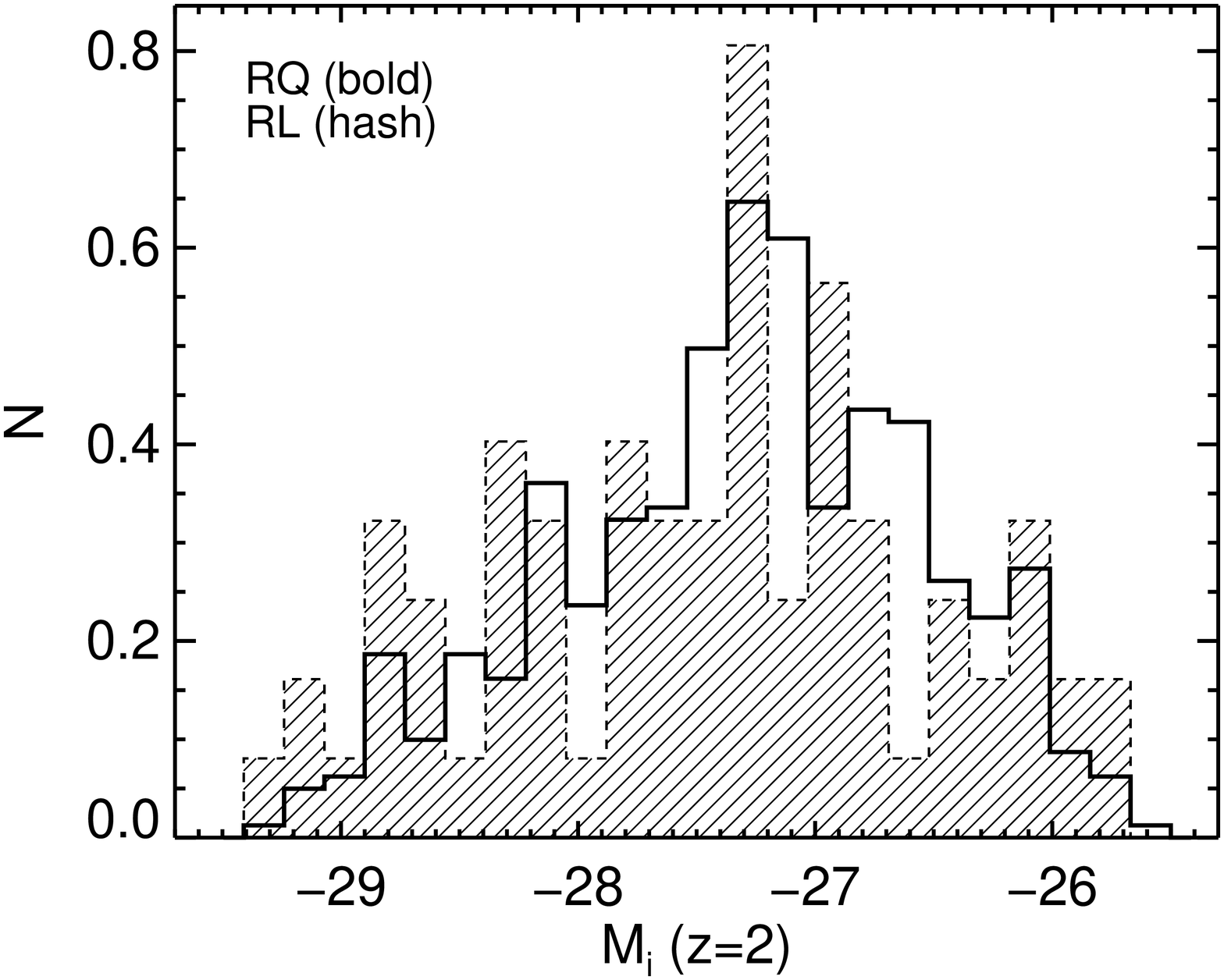}
   \vspace{0cm}
   \caption{The samples are selected to be well-matched in $z$ (top panel) and $i$-band magnitude, resulting in well-matched luminosities (plotted as $M_i$, $k$-corrected to $z=2$, bottom panel).  Histograms are normalized to an area of 1.}
   \label{fig:histo1}
\end{figure}

\begin{figure}
\centering
\hspace{0cm}
   \includegraphics[width=5.5cm]{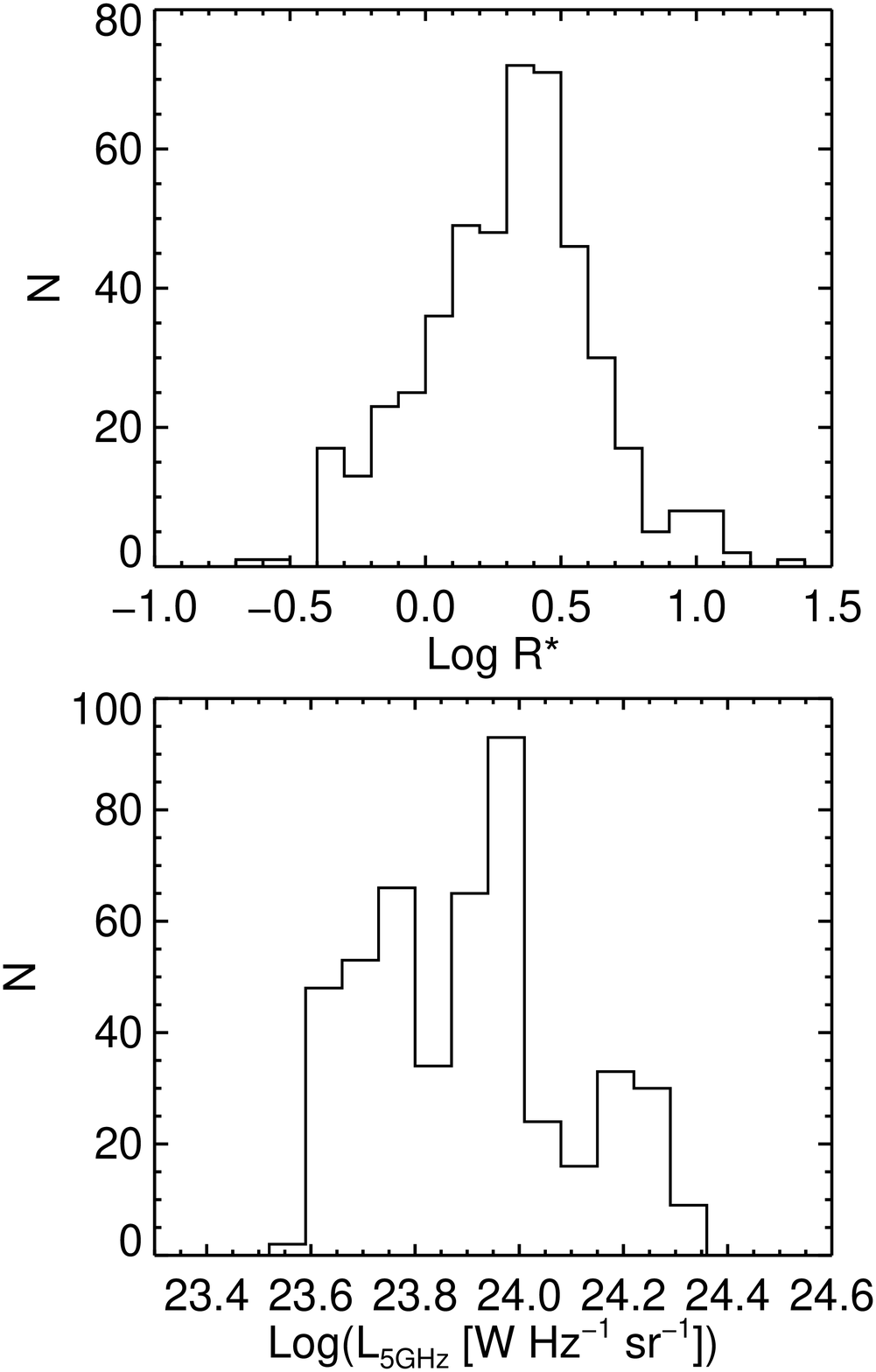}
   \vspace{0cm}
   \caption{The distributions of upper limits of the radio-loudness parameter $R^*$ (top panel) and the radio luminosity at 5 GHz (bottom panel) for the RQ sample, at the 1 mJy flux limit of the FIRST survey.  Using $R^*$, only 11 objects could be intrinsically radio loud, but using the radio luminosity none make the formal cutoff of $10^{25}$ W Hz$^{-1}$ sr$^{-1}$.\label{fig:limits}}
\end{figure}

\begin{figure*}
\centering
\hspace{0cm}
   \includegraphics[width=4.35cm]{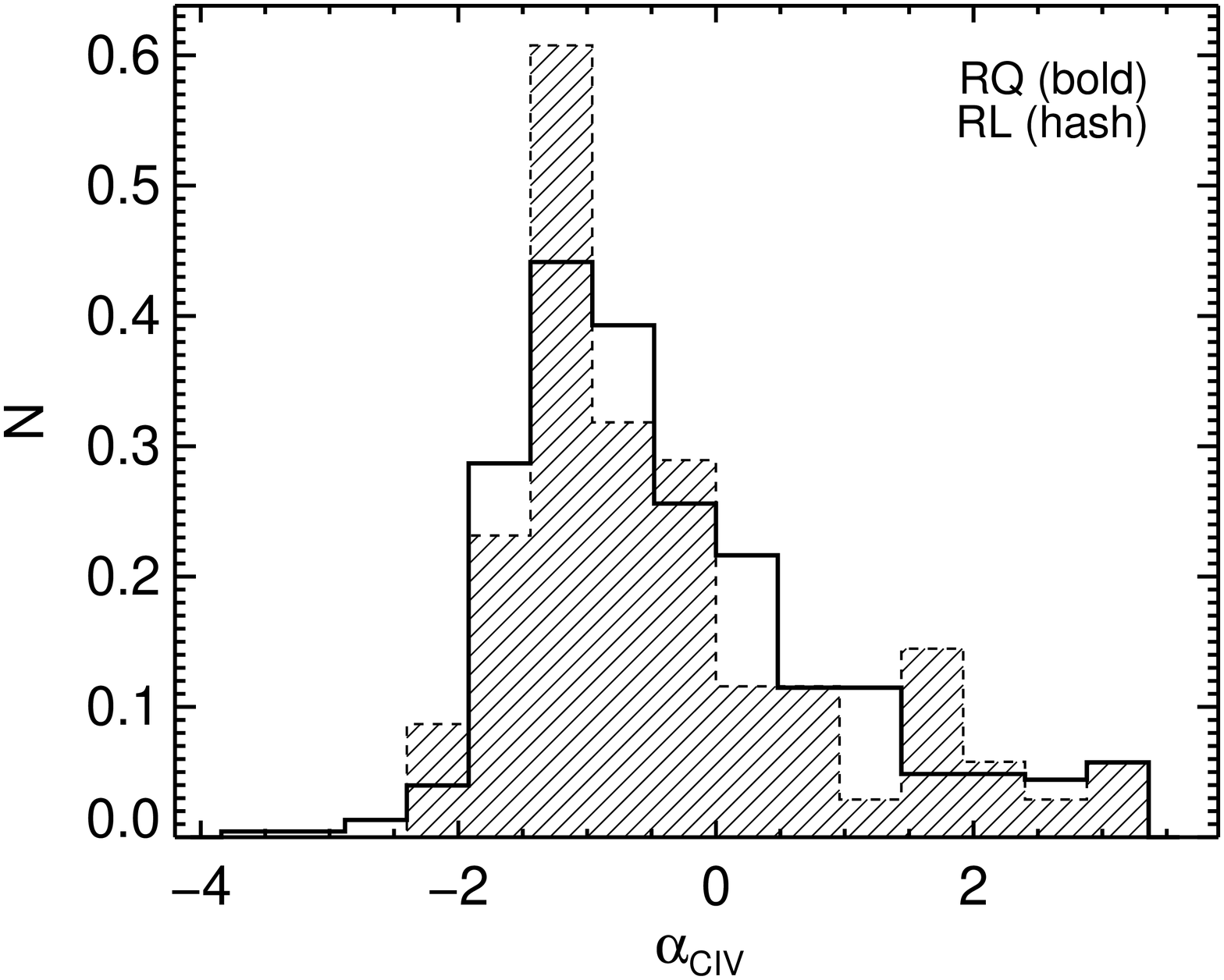}
   \includegraphics[width=4.35cm]{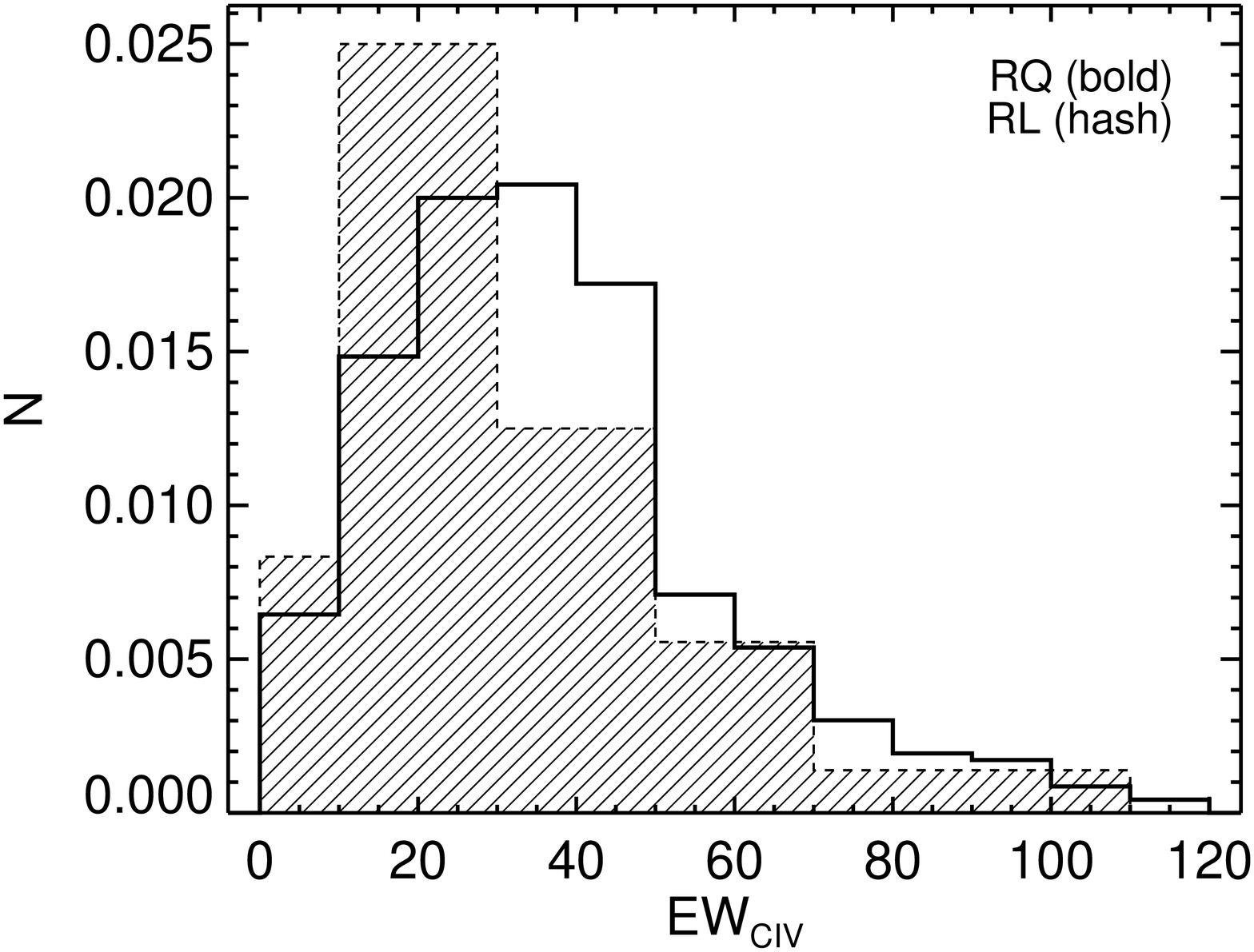}
   \includegraphics[width=4.35cm]{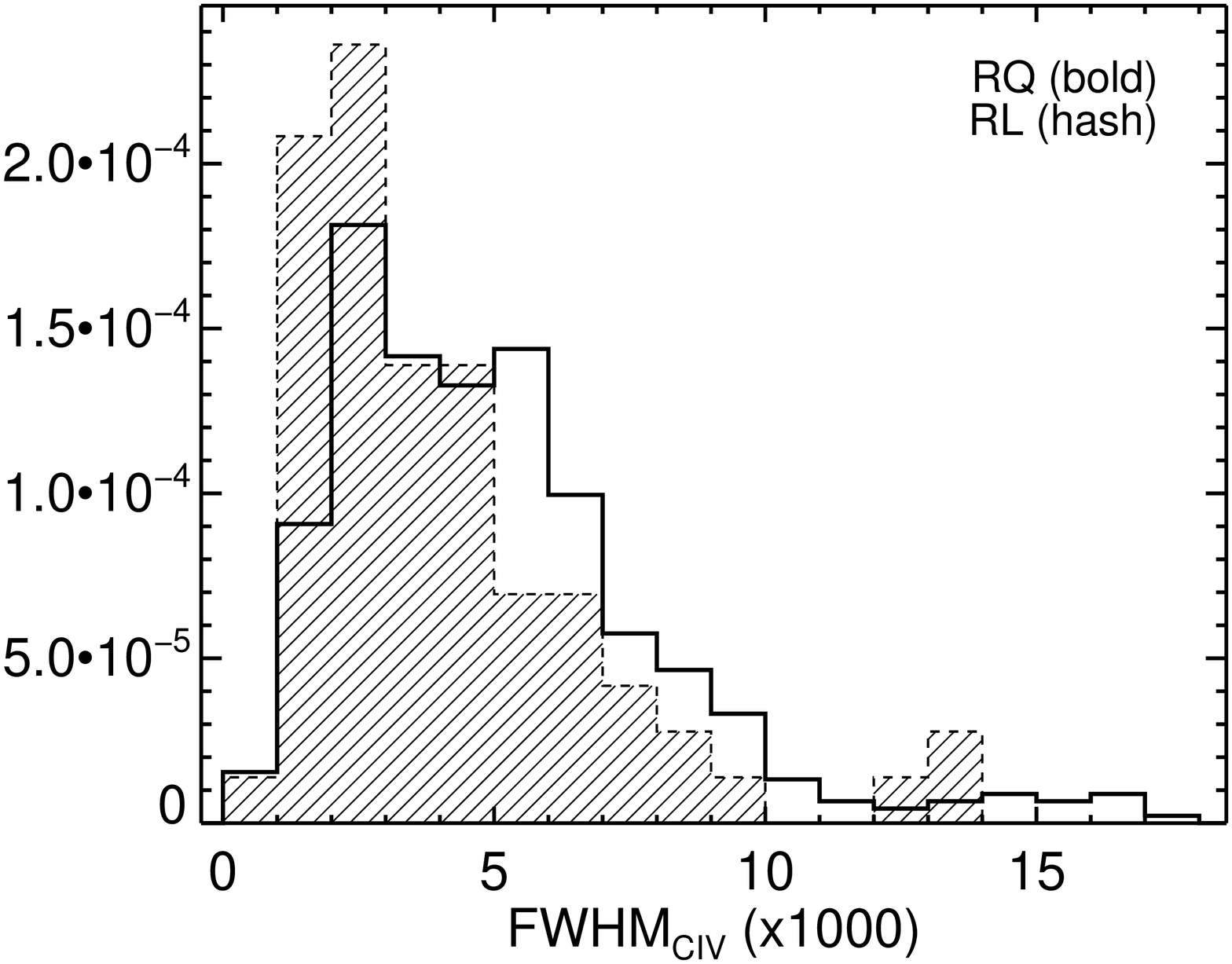}
   \includegraphics[width=4.35cm]{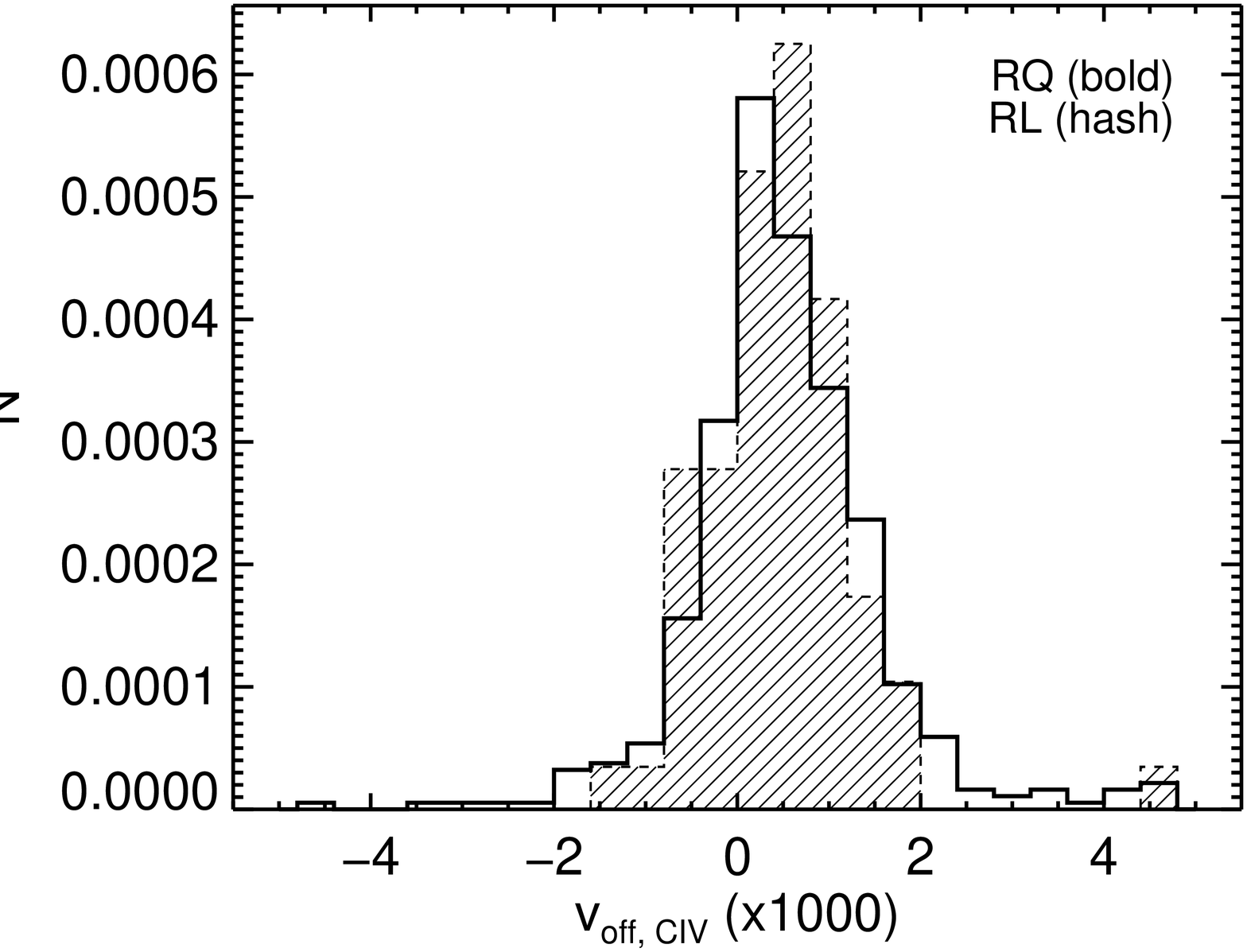}
   \includegraphics[width=4.35cm]{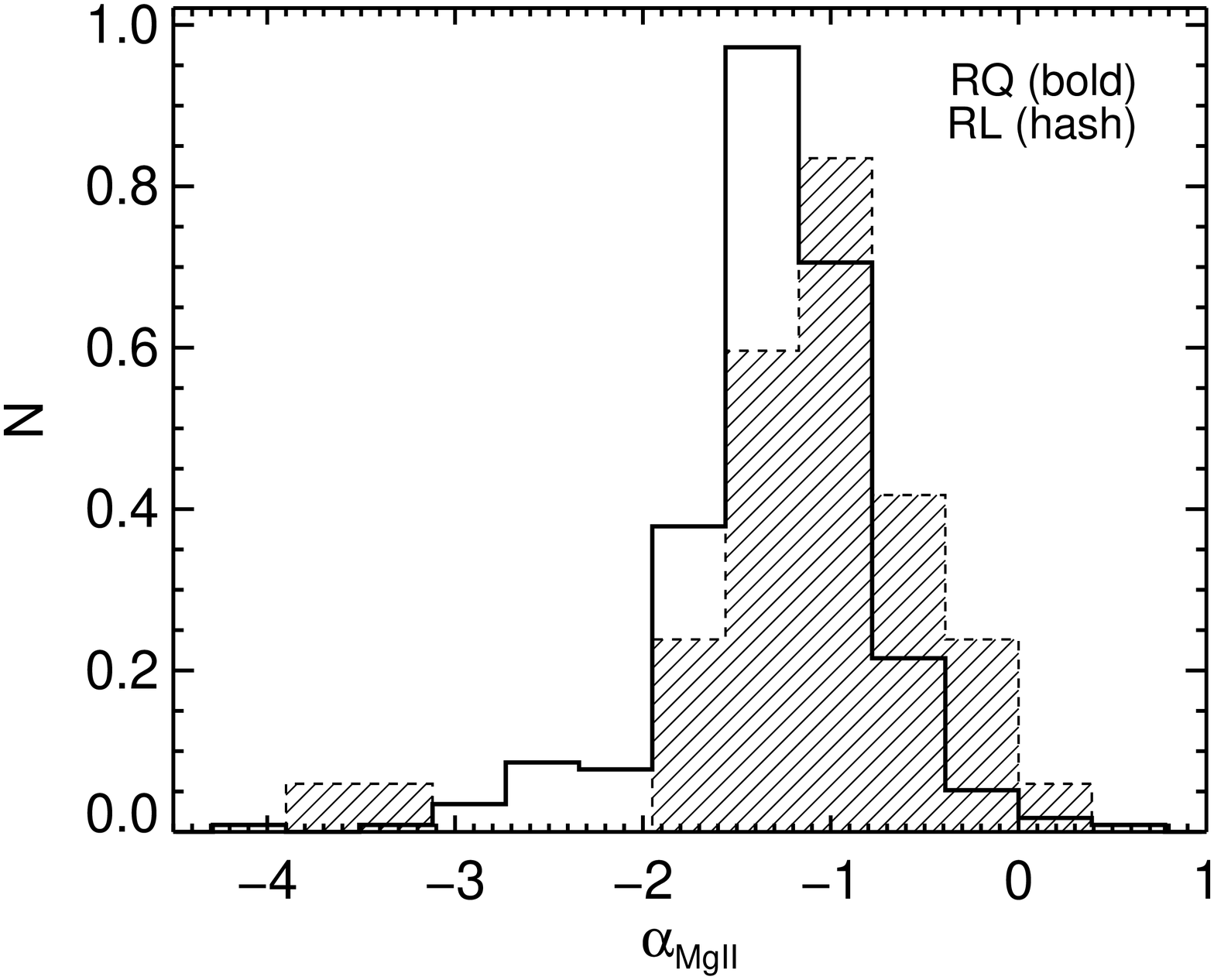}
   \includegraphics[width=4.35cm]{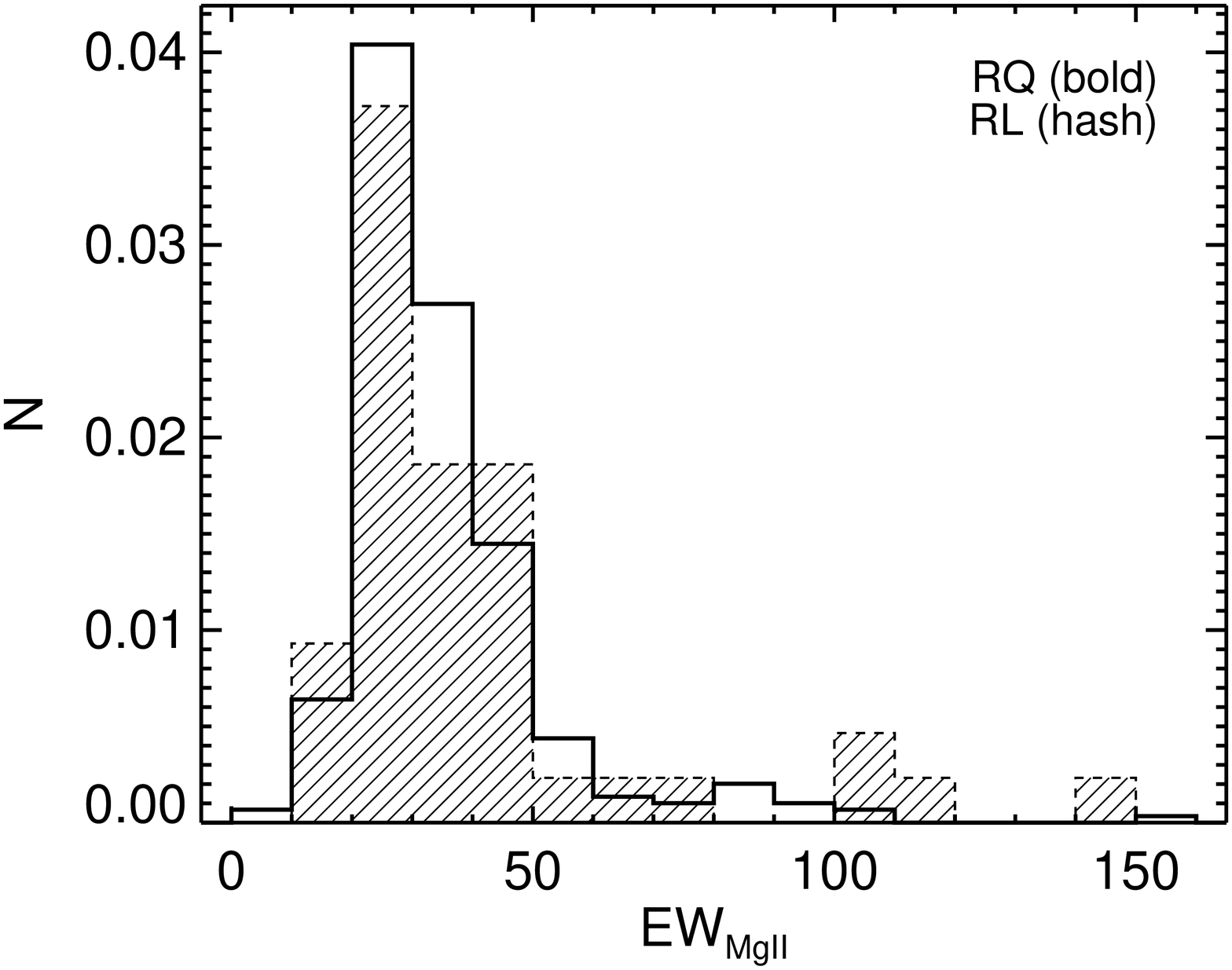}  
   \includegraphics[width=4.35cm]{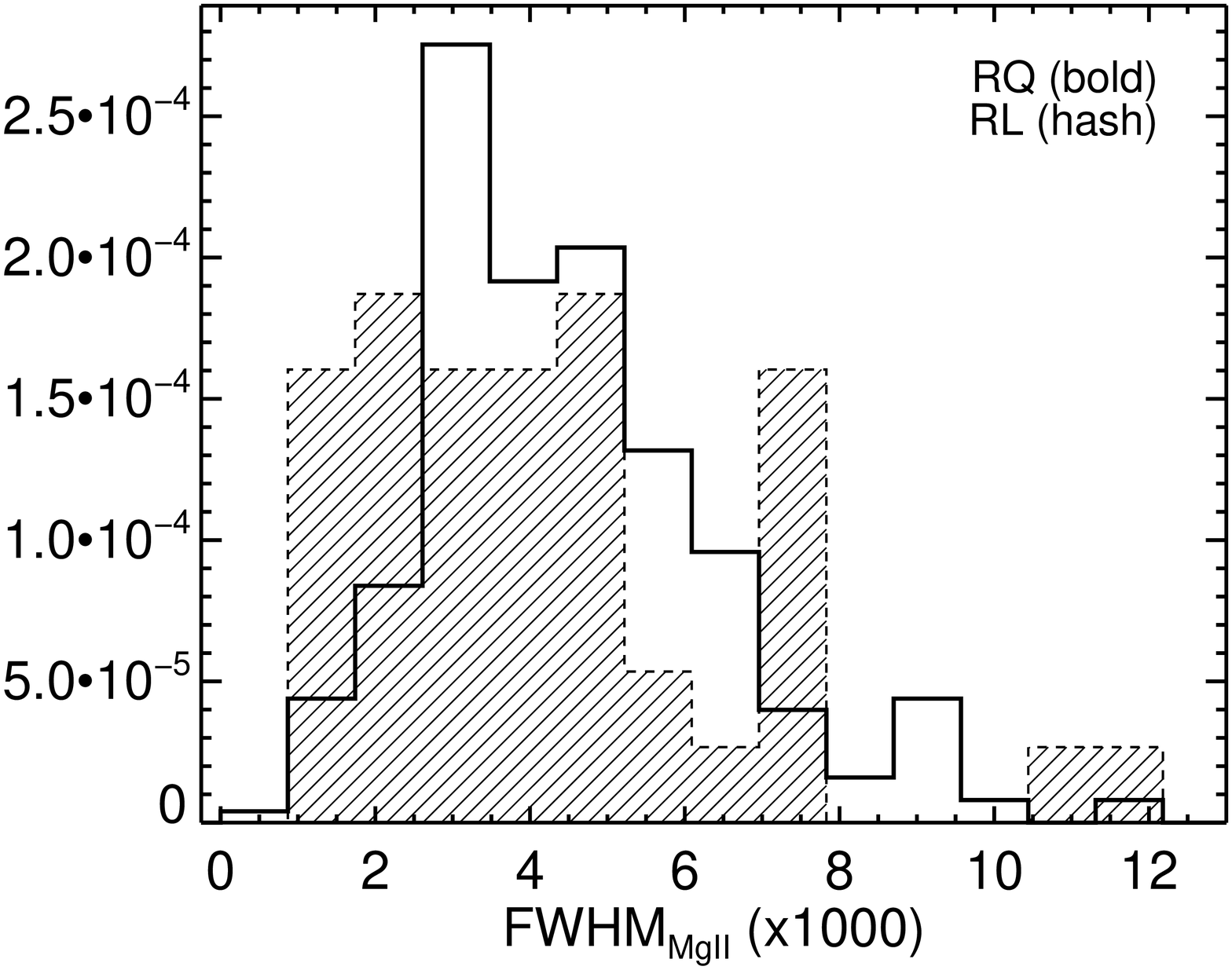}
   \includegraphics[width=4.35cm]{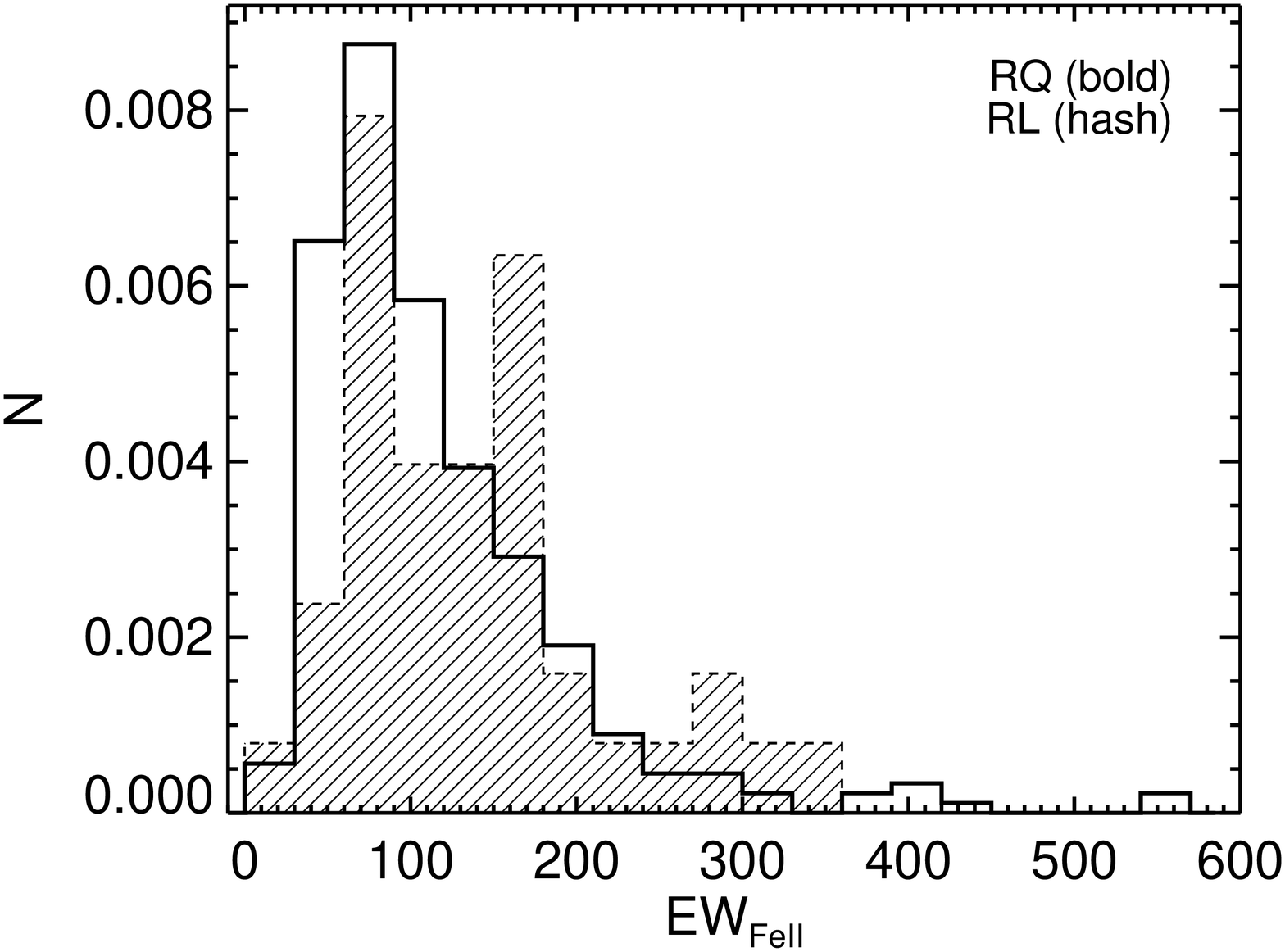}   
   \vspace{0cm}
  \caption{Comparison of the EW, FWHM, and continuum slope in the vicinity of the \CIV\ and \MgII\ emission lines.  Also shown are the velocity offset (relative to systemic) of \CIV, for a comparison with \citet{Richards11}, and the EW of \FeII\ emission in the region around \MgII.  Histograms are normalized to an area of 1.\label{fig:histo2}}
\end{figure*}

\begin{figure*}
\centering
\hspace{0cm}
   \includegraphics[width=5.5cm]{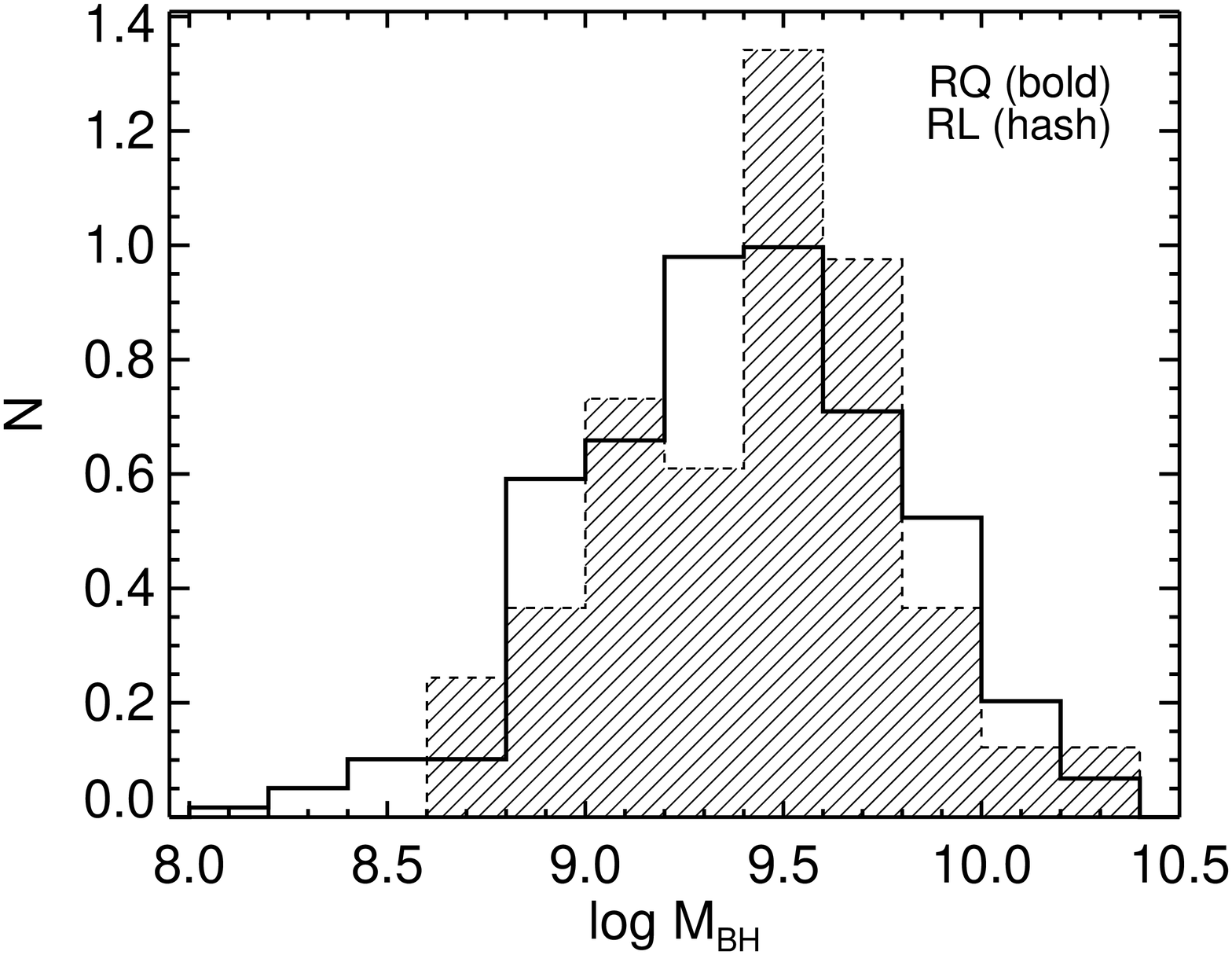}
   \includegraphics[width=5.5cm]{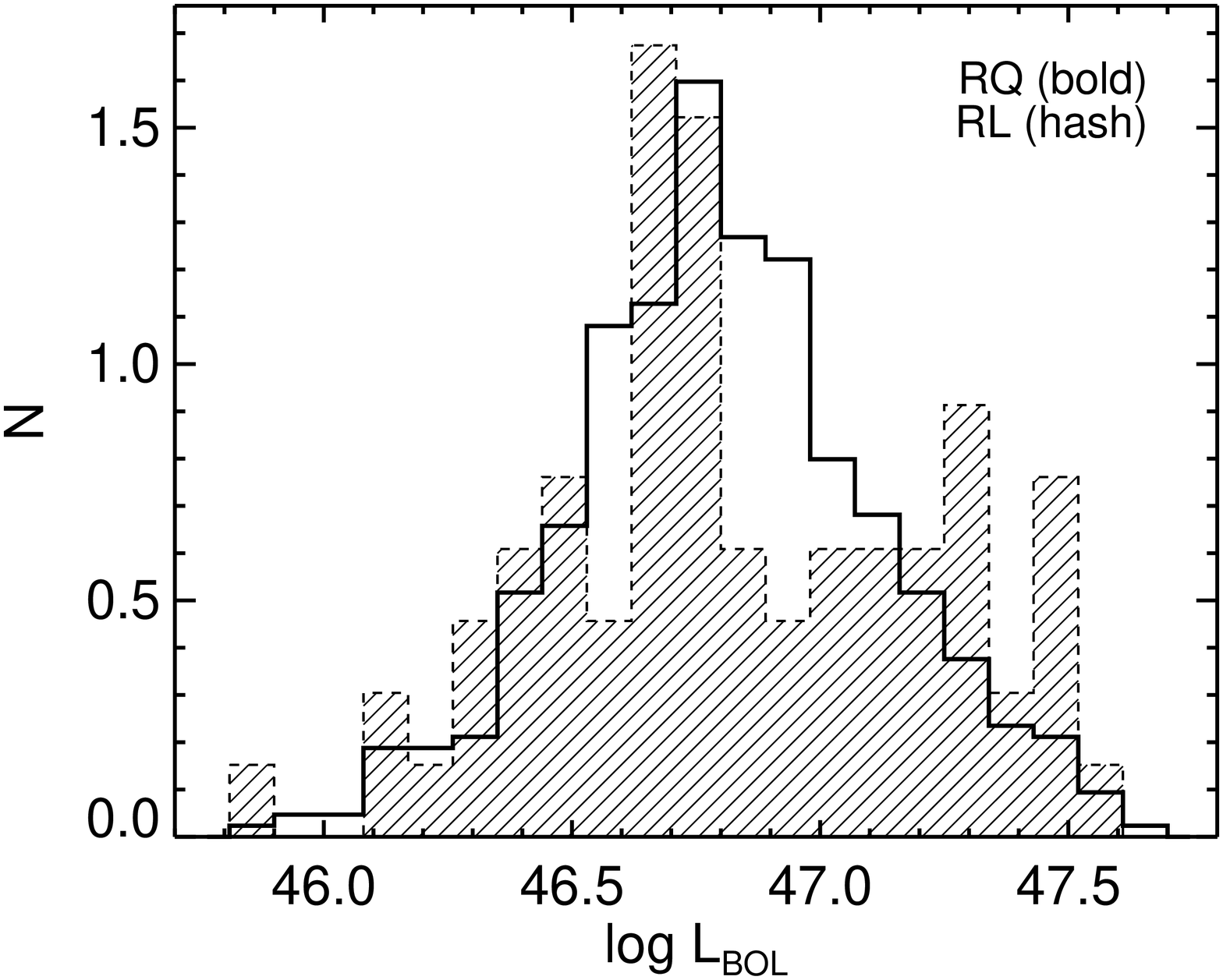}
   \includegraphics[width=5.5cm]{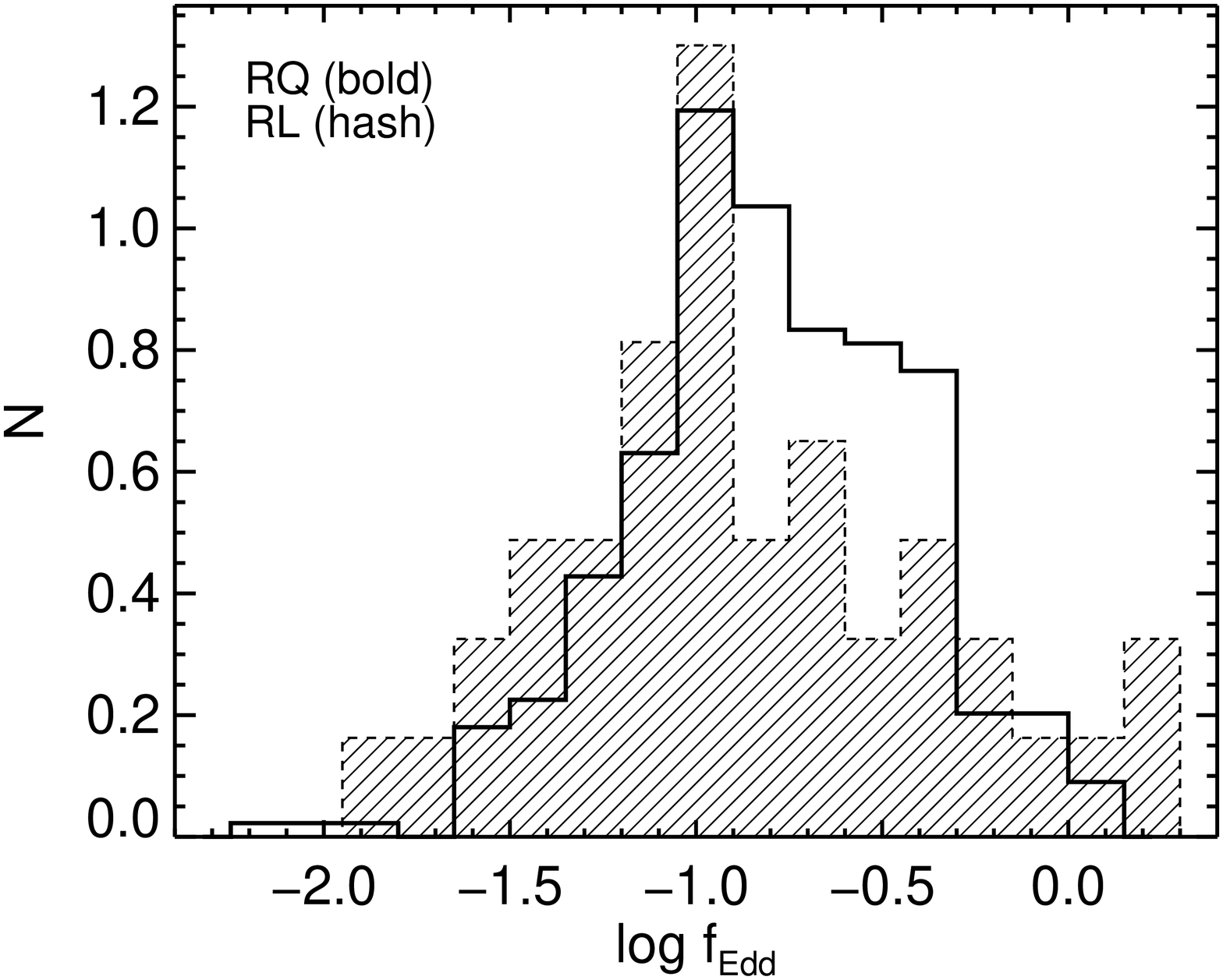} 
   \vspace{0cm}
  \caption{Comparison of some fundamental physical properties derived from the spectral fits --- black hole mass ($M_{\textrm{BH}}$; left), bolometric luminosity ($L_{\textrm{bol}}$; center), and Eddington fraction ($f_{\textrm{Edd}}$; right).  Histograms are normalized to an area of 1).}
  \label{fig:histo3}
\end{figure*}

\begin{figure*}
\centering
\hspace{0cm}
   \includegraphics[width=5.5cm]{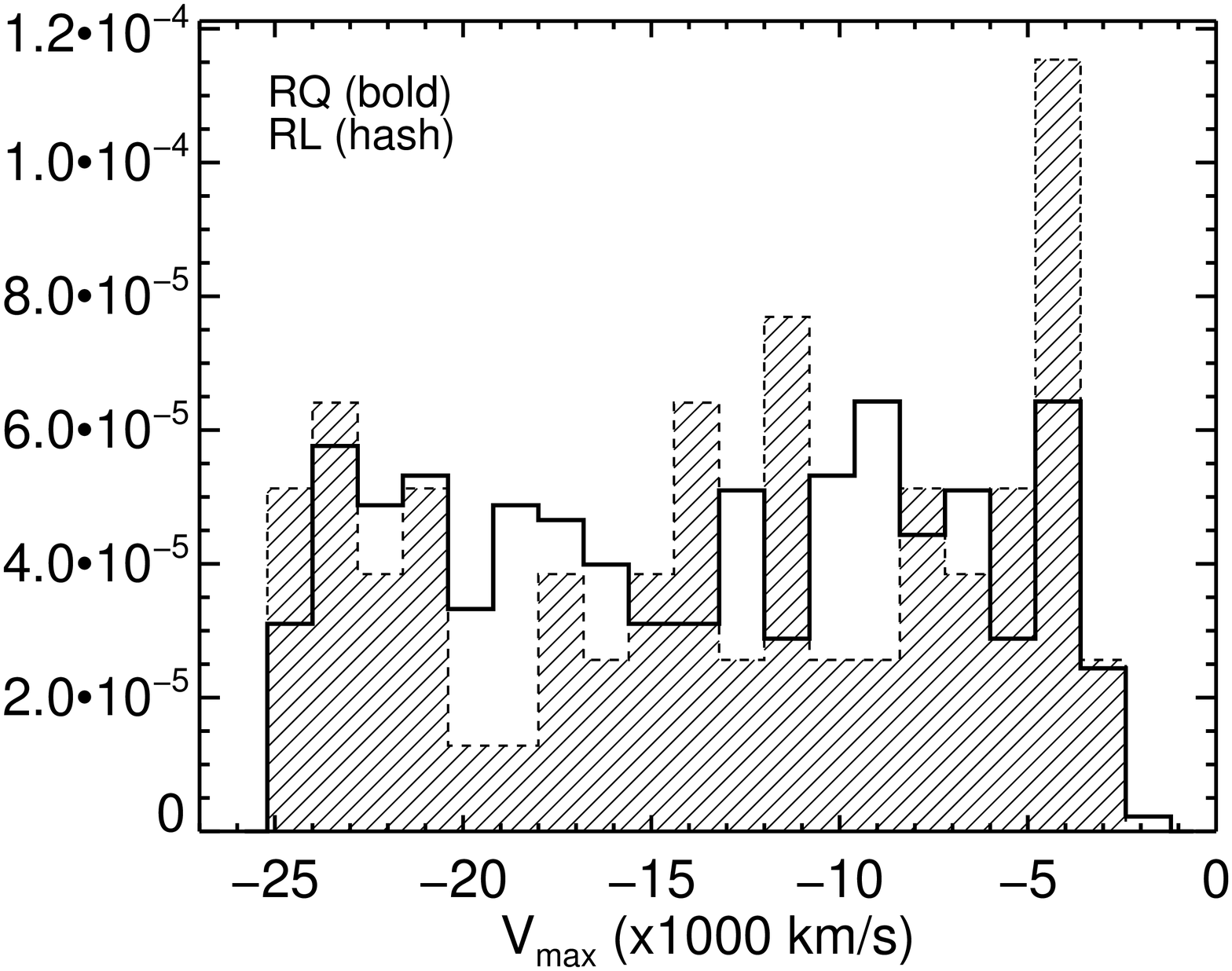}
   \includegraphics[width=5.5cm]{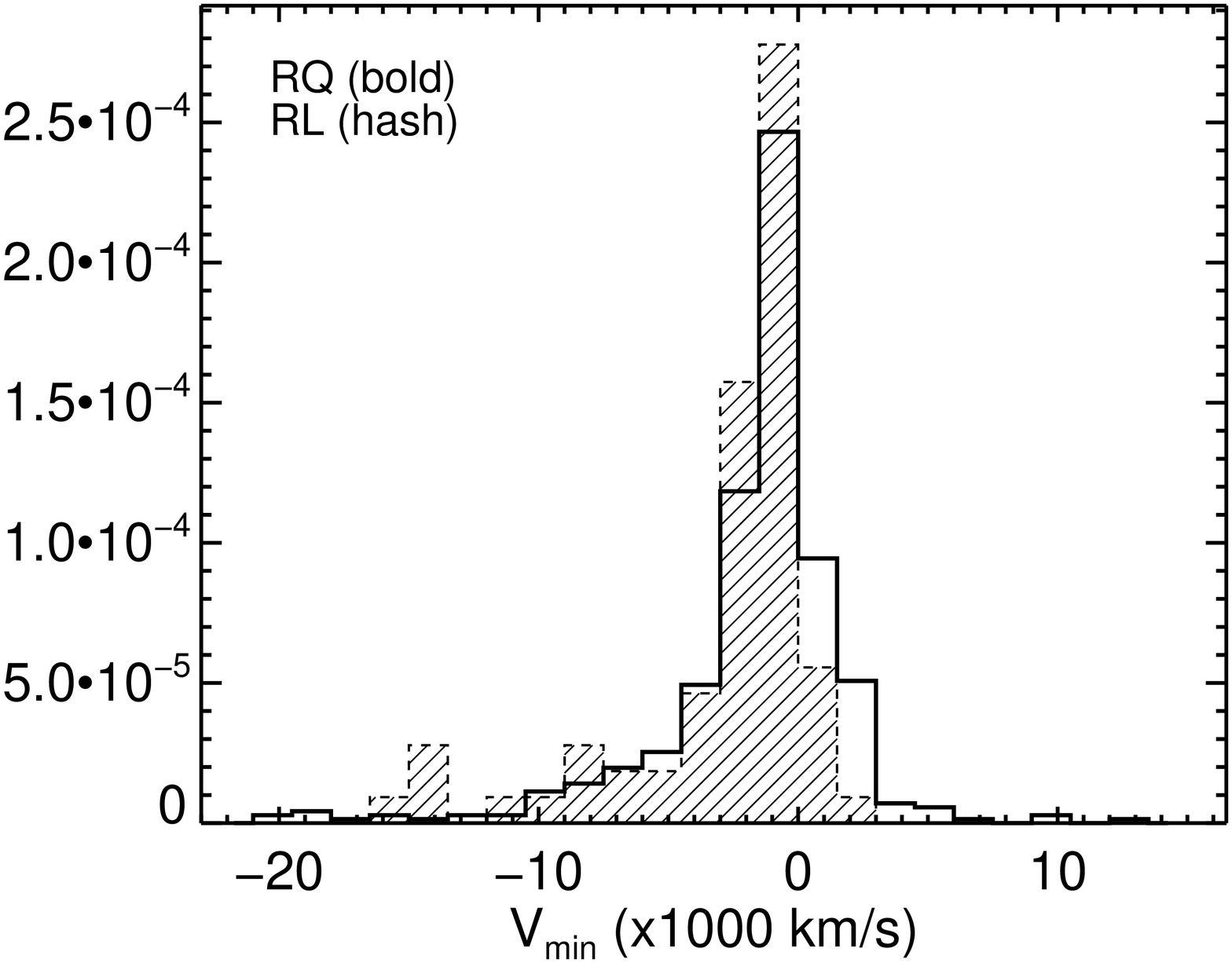}
   \includegraphics[width=5.5cm]{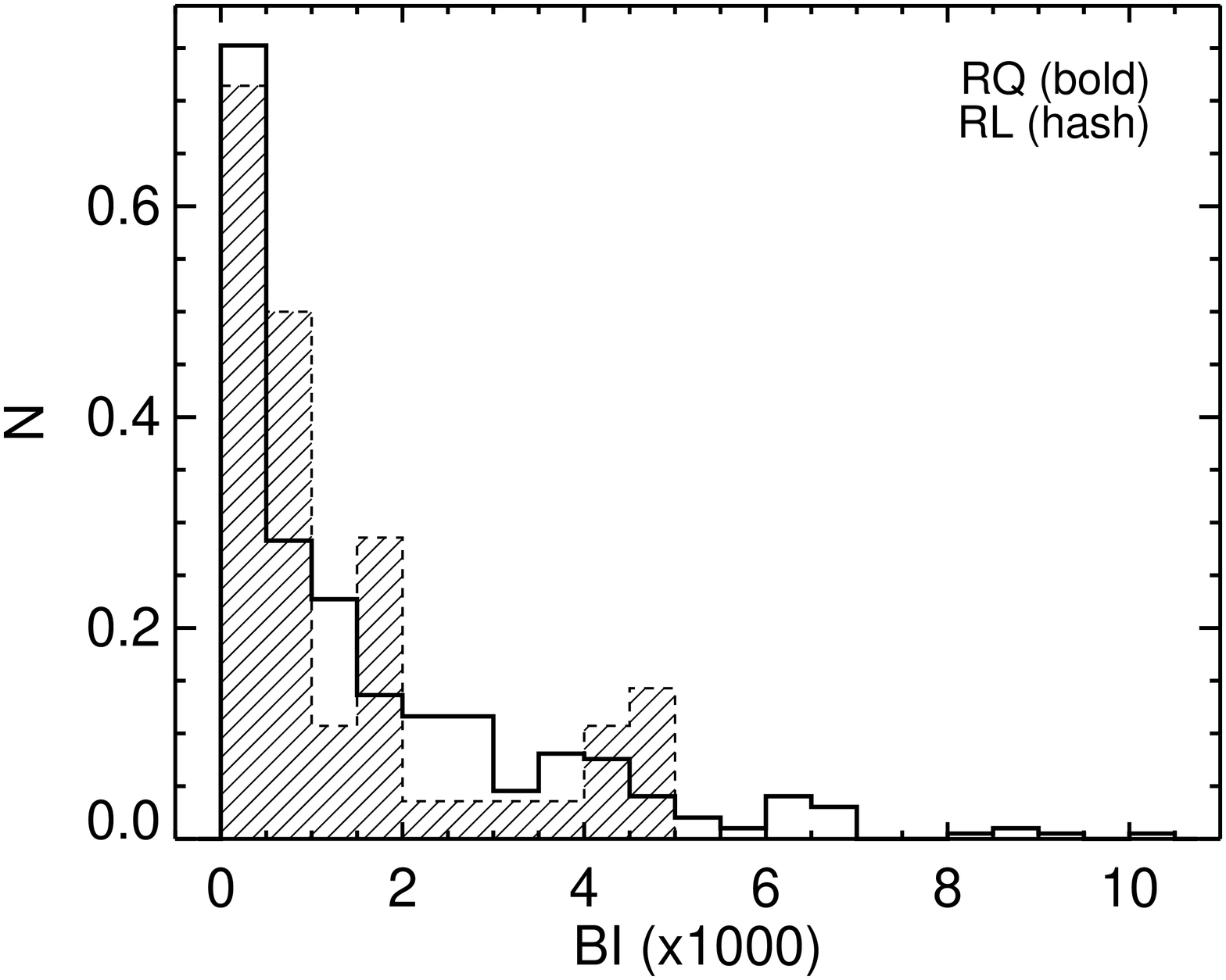}
   \includegraphics[width=5.5cm]{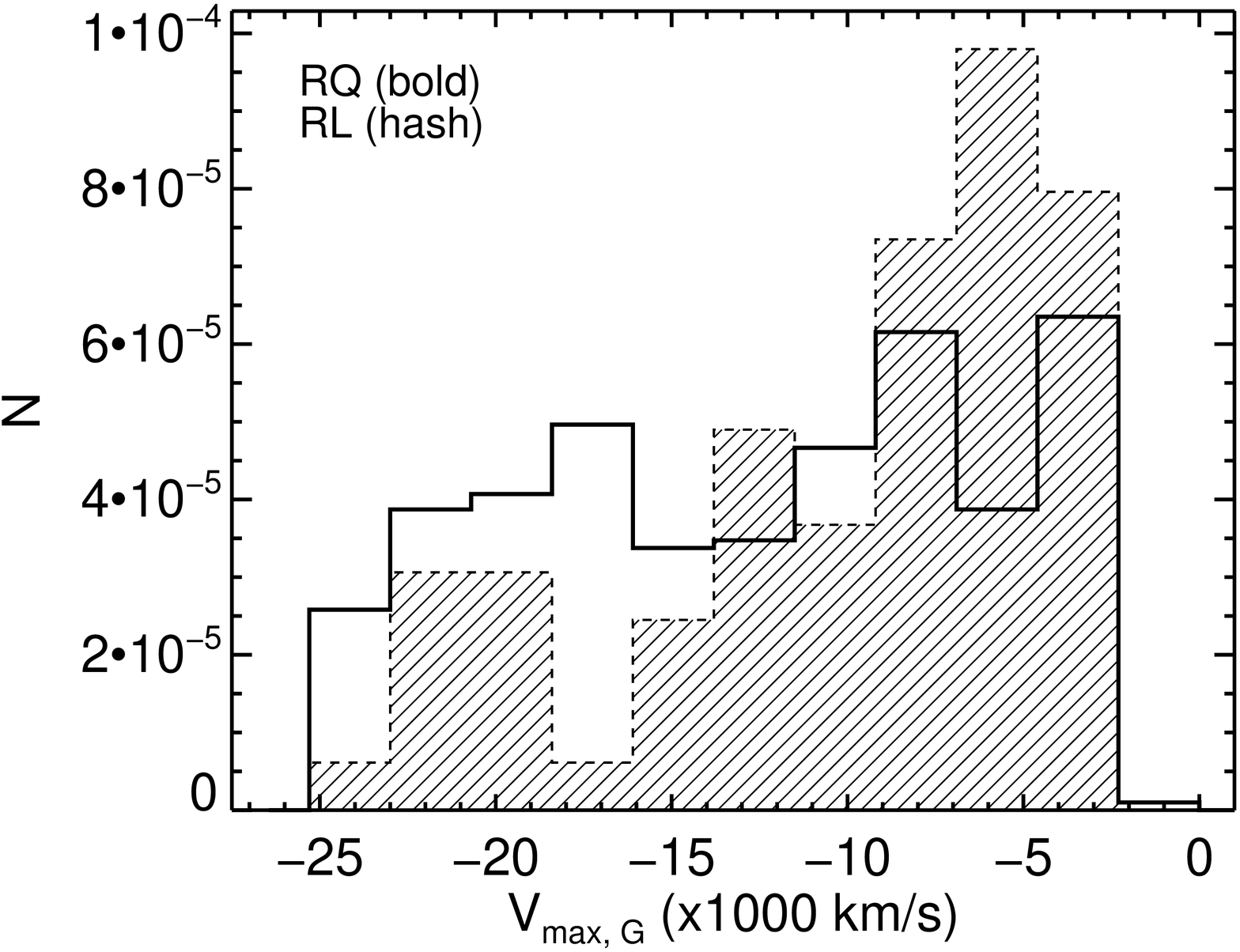}
   \includegraphics[width=5.5cm]{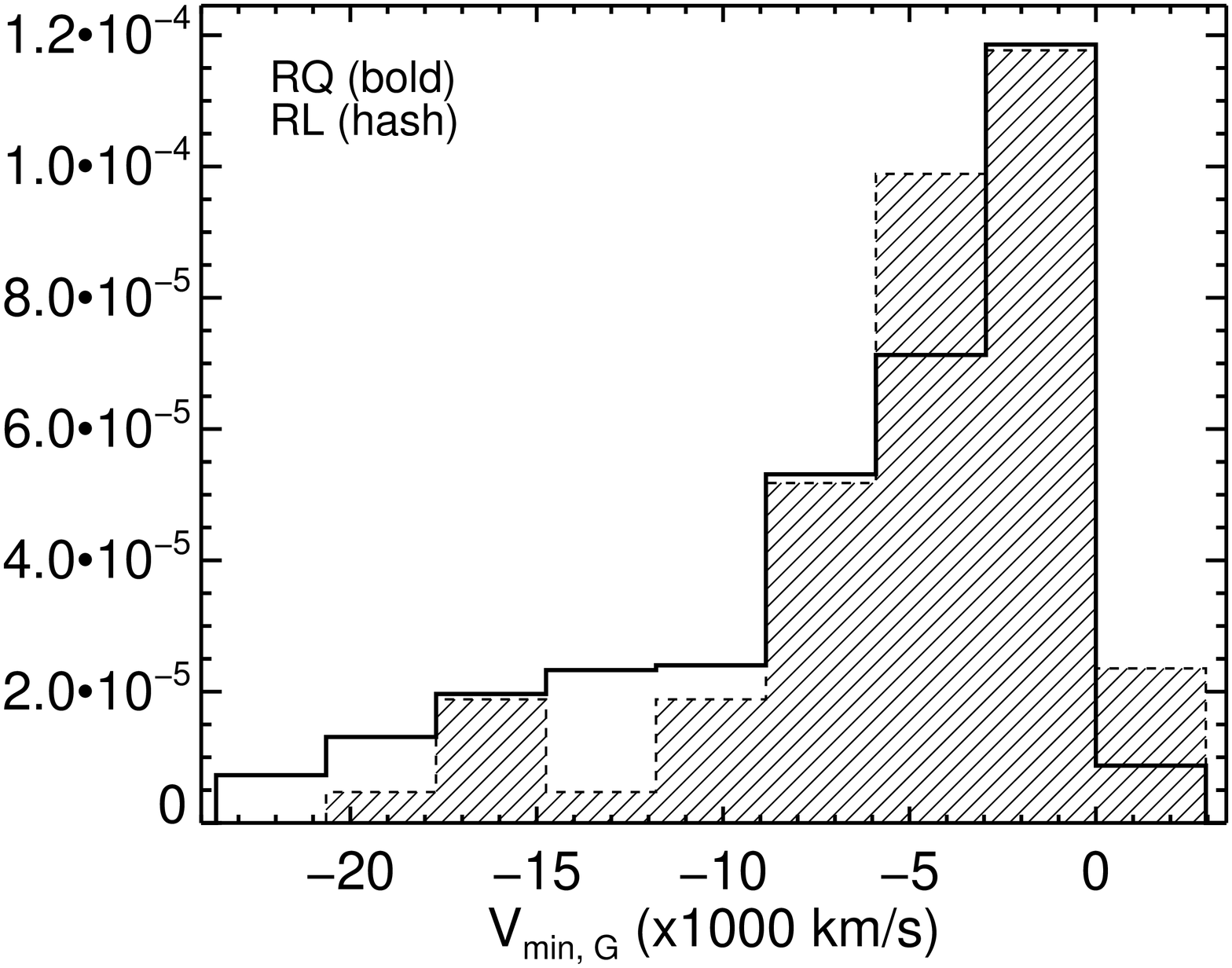}
   \includegraphics[width=5.5cm]{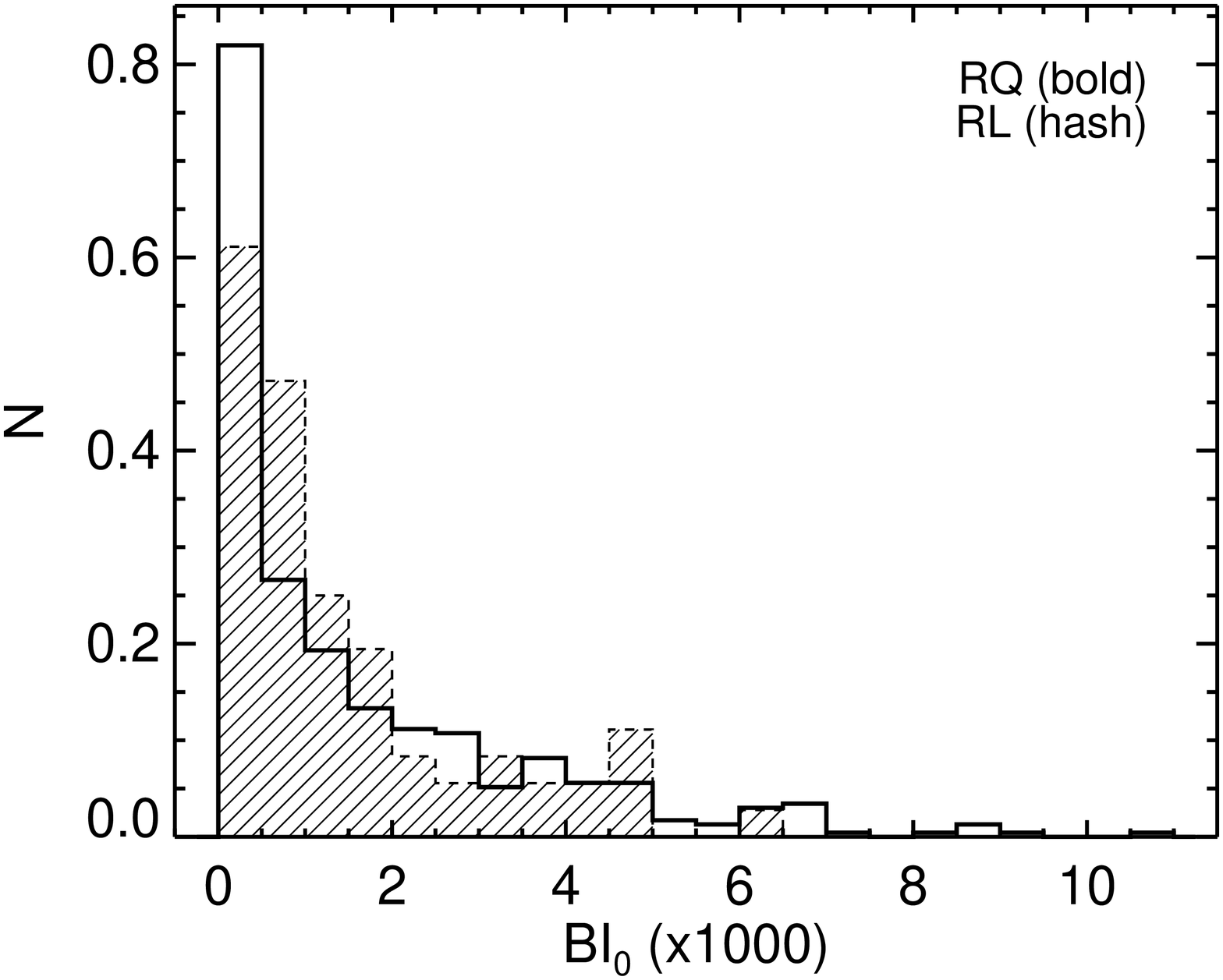}    
   \vspace{0cm}
  \caption{Comparison of several BAL parameters, including the minimum velocity as measured here and by G09 (top and bottom left, respectively), the maximum velocity measured here and by G09 (top and bottom center, respectively), the traditional balnicity index ($BI$, top right), and the modified balnicity index ($BI_0$, bottom right).  Histograms are normalized to an area of 1.}
  \label{fig:histo4}
\end{figure*}

\section{ANALYSIS \& RESULTS}
The key purpose of this work is to perform a detailed comparison of the rest-frame UV spectral properties of RL and RQ BAL quasars. We use measurements from the catalog of G09 and \citet[hereafter S11]{ShenCatalog11}, which are built from the SDSS Data Release 5 and 7, respectively.  We refer the reader to G09 and S11 for full details of their measurements and catalogs, but provide a brief summary here. 

G09 provides measurements of several BAL properties, such as minimum and maximum absorption velocities, as well as traditional and modified balnicity indices ($BI$ and $BI_0$, respectively).  S11 provides extensive emission line and continuum spectral fits, along with physical parameters derived from these fits. \citet{DiPompeo12} made independent fits to the SDSS spectra of our RL sample and checked the robustness of the values in the G09 and S11 catalogs. Though there was some scatter, as expected, in general the results agreed fairly well \citep[][figures 3, 4, and 5]{DiPompeo12}. 

One exception to this agreement is the measurement of the absorption velocities, due to slight differences in definitions.  G09 measured velocities based on their continuum fits, and have strict requirements on what qualifies as absorption. \citet{DiPompeo12} used the more subjective visual inspection of the absorption lines, which allows the identification of more subtle absorption features and is more sensitive to absorption on top of the emission lines. While in many cases these methods agree well, there are some cases where the G09 and \citet{DiPompeo12} velocity measurements can differ significantly. Because of this, we include analysis of absorption velocities using our own measurements from visual inspection as well as values from the G09 catalog. 

From the S11 catalog we adopt measurements of the \MgII\ and \CIV\ emission lines, the continuum slope measured around these lines ($\alpha_{CIV}$ and $\alpha_{MgII}$), and \FeII\ emission around the \MgII\ emission line, as well as several derived physical properties. S11 uses up to three Gaussians to fit the \MgII\  and \CIV\  emission lines. A local power-law was fit to the continuum ($f_\lambda = A\lambda^{\alpha_\lambda}$ ), in addition to an iron template \citep{Vestergaard01}, in windows on either side of the emission lines. 

\CIV\  emission lines are difficult to fit in BAL quasar spectra, particularly in cases where absorption removes significant portions of the emission lines (i.e. objects with small $v_{min}$).  There are several techniques employed in various studies to avoid these complications \citep{Weymann91, Gibson08, Allen11}, but the simple fact remains that it is difficult to accurately quantify the intrinsic \CIV\  emission properties in BAL quasars. Thus, to provide an accurate comparison, we avoid adopting physical parameters that were derived from the \CIV\ emission line fits --- black hole (BH) masses and Eddington ratios are adopted using fits and calibrations for \MgII\ only. Specifically, S11 used the following relationship to calculate the \MgII\  virial black hole mass:

\begin{equation}
\begin{multlined}
\log \left( \frac{M_{\textrm{BH}}}{M_{\sun}} \right) = 0.740 + 0.62 \log \left( \frac{\lambda L_\lambda}{10^{44} \ \textrm{erg} \ \textrm{s}^{-1}} \right) \\
 + 2 \log \left( \frac{\textrm{FWHM}}{\textrm{km} \ \textrm{s}^{-1}} \right)
\end{multlined}
\end{equation}
	
\noindent where $M_{\rm{BH}}$ is the \MgII\  virial black hole mass, $\lambda L_\lambda$ is the monochromatic luminosity (at 3000 \AA\ for the \MgII\  BH mass calibration), and FWHM is the full-width at half-maximum of the combined Gaussian fits (with a narrow component subtraction) to the \MgII\  line.  We then use the relationship:

\begin{equation}
L_{\textrm{Edd}} \hspace{0.1cm} \textrm{[ergs/s]} = 1.51 \times 10^{38} \left( \frac{M_{\textrm{BH}}}{M_{\sun}} \right)
\end{equation}

\noindent \citep[e.g., see][]{Peterson03} to calculate the Eddington luminosity and Eddington fraction ($f_{\textrm{Edd}} = L_{\textrm{bol}}/L_{\textrm{Edd}}$, where $L_{bol}$ is adopted from S11) based on just the \MgII\  BH masses.

We use the Kolmogorov-Smirnov (KS) test to compare several properties of RL and RQ quasars. Table 2 summarizes these comparisons.  Column 1 lists the parameter being compared (described in the caption where necessary), columns 2 and 3 give the number of RL and RQ objects with a measurement available, respectively, columns 4 through 9 give some basic statistics of each parameter for each subsample, and columns 10 and 11 give the KS test statistic and associated probability that the distributions are drawn from the same parent population. 

We choose a cutoff for ``significant'' differences in distributions of $P=0.05$, or roughly a 2$\sigma$ difference. Distributions with P-values lower than this are possibly from different parent populations, and are highlighted in bold in the table.

To test the effect of the large difference in sample sizes, we use Monte-Carlo resampling. We randomly sample the RQ population to the same sample size of the RL population (using $N_{\textrm{RL}}$ appropriate for each parameter), and compare these with the KS test. We repeat this process $10^4$ times, and record the percentage of trials for which $P < 0.05$. This percentage is listed in column 12 of table 2. In general, this percentage is consistent with what we expect from the KS tests using the full samples, and there is no reason to believe that the differences or similarities found are driven by differences in the sample sizes.

The KS tests indicate that in general, the properties of RL and RQ BAL quasars are quite similar.  The parameters that may have significant ($P_{\textrm{KS}} < 0.05$) differences are the continuum slope around \MgII\ ($\alpha_{\textrm{MgII}}$), the strength of \FeII\ emission (as measured by the equivalent width (EW)), the EW and FWHM of \CIV, the maximum absorption velocity as measured by G09 ($v_{\textrm{min, G}}$), and the minimum absorption velocity measured here (and very nearly as measured by G09 as well, at $P_{\textrm{KS}} = 0.064$, though in the opposite direction).  Several of these differences are interrelated, and will be discussed further in the next section.

As an additional statistical test, we perform Wilcoxon Rank Sum (RS) tests on each parameter, which are more sensitive to differences in the means of the distribution. Our results, summarized in column 14 of table 2, confirm the results from the KS tests.

Finally, we perform a $\chi^2$ test on each parameter, which is a slightly less conservative way to compare the distributions, and can possibly identify more subtle differences between the subclasses.  To do this, for a given parameter we divide the smaller RL sample into bins of varying width, such that each bin contains 5 quasars (except for the last bins, which may contain more than 5 due to rounding). We then apply this binning to the RQ sample, normalizing each bin by dividing by the total number of RQ objects and multiplying by the total number of RL objects (columns 3 and 2 of Table 2, respectively).  If a bin contains fewer than 5 RQ BAL quasars, then adjacent bins are merged (for both samples) until there are at least 5 objects per bin.  The $\chi^2$ statistic is calculated by using the number of RL quasars in each bin as the ``observed'' frequency, and the number of RQ quasars in each as the ``model'' frequency, and summing over each bin.  The associated probability that the distributions are the same is calculated from $\chi^2$ --- these probabilities are given in column 13 of Table 2.  Values of $P_{\chi^2} < 0.05$ are also highlighted in bold, and generally confirm the results from the KS and RS tests.

The number of LoBALs in each sample is not insignificant (on the order of $\sim$10\%), and these objects can potentially skew the results, particularly for comparisons of \MgII\ properties where additional absorption can affect the emission line and surrounding continuum fits.  These fitting complications can propagate into the comparisons of physical parameters based on \MgII\ parameters, such as the BH masses.  As a check, we also compare all of the properties excluding the LoBALs from the samples.  We find that our results do not change significantly.

In addition to the statistical tests above, we also create composite spectra of the RL BAL and RQ BAL samples for a visual comparison of the mean spectra.  To create the composite, we normalize each spectrum in the flux range 2000-2050 \AA\ (one of the few continuum regions that is available in all of the spectra and is relatively free of emission features) and average them using a 3$\sigma$ rejection.  The composites are shown in Figure 5 (top panel), along with the RMS residual spectrum of each to illustrate the amount of variation within a sample (middle panel), and finally a residual spectrum of the RL composite divided by the RQ composite is shown in the bottom panel.

\begin{table*}
\centering
  \caption{RL BAL and RQ BAL property statistics and comparisons.}
  \label{table:comparison}

  \begin{tabular}{lccccccccccccr}
    \hline
Property & $N_{\textrm{RL}}$ &  $N_{\textrm{RQ}}$  &  $\mu_{\textrm{RL}}$ &  $\mu_{\textrm{RQ}}$  &  $Med_{\textrm{RL}}$   &  $Med_{\textrm{RQ}}$   &  $\sigma_{\textrm{RL}}$ &  $\sigma_{\textrm{RQ}}$  & $D_{\textrm{KS}}$ & $P_{\textrm{KS}}$ & $\%$ & $P_{\chi^2}$ & $P_{\textrm{RS}}$\\
\hline
$z$&     73&    473&      2.20&      2.16&      2.07&      2.08&      0.49&      0.46&     0.081&     0.779&      0.56&      0.954& 0.276 \\
$\mathrm{M_i (z=2)}$   & 73 & 473 & -27.46  & -27.31  & -27.37 & -27.25 & 0.91    & 0.77    & 0.132 & 0.202 & 3.54  & 0.115 & 0.080\\
$\mathrm{\alpha_{CIV}}$&     72&    472&     -0.37&     -0.37&     -0.67&     -0.66&      1.27&      1.25&     0.080&     0.808&      1.23&      0.718& 0.481\\
$\mathrm{\alpha_{MgII}}$&     43&    298&     -1.02&     -1.31&     -0.99&     -1.27&      0.95&      0.55&     0.297&     \textbf{0.002}&     91.22&     \textbf{0.0001}& \textbf{0.002}\\
$\mathrm{EW_{FeII}}$&     42&    297&    165.04&    115.21&    135.48&     95.78&    141.59&     79.71&     0.248&     \textbf{0.018}&     56.40&      \textbf{0.025}& \textbf{0.003}\\
$\mathrm{EW_{CIV}}$   & 72 & 465 & 31.48   & 38.73   & 25.94   & 34.78   & 21.15   & 30.85   & 0.209 & \textbf{0.007} & 57.22 & 0.182 & \textbf{0.004}\\
$\mathrm{FWHM_{CIV}}$ & 72 & 452 & 4118.46 & 5056.64 & 3518.10 & 4603.55 & 2736.63 & 2979.86 & 0.224 & \textbf{0.003} & 63.12 & \textbf{0.034} & \textbf{0.001}\\
$v_{\textrm{off, CIV}}$ &     72&    465&    450.39&    504.78&    487.61&    423.44&    825.16&   1040.63&     0.080&     0.800&      0.81&      0.569& 0.344\\
$\mathrm{EW_{MgII}}$&     43&    297&     41.18&     35.98&     31.20&     30.64&     28.72&     23.60&     0.100&     0.830&      0.11&      0.677& 0.332\\
$\mathrm{FWHM_{MgII}}$&     43&    288&   4248.88&   4594.09&   4038.14&   4301.13&   2448.63&   2111.37&     0.206&     0.072&     10.75&      0.129& 0.075\\
$v_{\textrm{off, MgII}}$&     41&    297&    248.78&     92.89&    477.45&    156.95&   1195.88&    895.50&     0.220&     0.052&     41.07&      0.058& \textbf{0.011}\\
$\log f_{\textrm{Edd}}$&     41&    296&     -0.86&     -0.80&     -0.92&     -0.80&      0.51&      0.38&     0.180&     0.174&     9.15&      0.436& 0.124\\
$\log\ M_{\textrm{BH}}$&     41&    296&      9.47&      9.39&      9.52&      9.40&      0.41&      0.40&     0.151&     0.353&      4.04&      0.371& 0.178\\
$\log\ L_{\textrm{bol}}$&     73&    473&     46.84&     46.81&     46.76&     46.79&      0.39&      0.31&     0.125&     0.260&      2.30&     \textbf{0.014}& 0.388\\
$\mathrm{log\ L_{\textrm{3000}}}$&     43&    298&     46.11&     46.08&     46.09&     46.07&      0.36&      0.28&     0.130&     0.520&      1.49&  0.267& 0.288\\
$\mathrm{log\ L_{\textrm{1350}}}$&     72&    472&     46.19&     46.20&     46.14&     46.20&      0.41&      0.32&     0.129&     0.229&      8.75&  \textbf{0.003}& 0.282\\
$v_{\textrm{min, G}}$&     72&    466&  -4630.88&  -6351.01&  -3316.00&  -4641.00&   4473.51&   5630.05&     0.164&     0.064&     23.75&     \textbf{0.048}& \textbf{0.009}\\
$v_{\textrm{max, G}}$&     71&    438&  -9970.83& -12635.78&  -8509.00& -11942.00&   5848.01&   6524.20&     0.234&     \textbf{0.002}&     74.38&     \textbf{0.001}& \textbf{0.001}\\
$v_{\textrm{min}}$&     72&    473&  -2808.66&  -1697.58&  -1349.49&  -1036.00&   3885.99&   3695.46&     0.170&     \textbf{0.048}&     27.94&      0.167& \textbf{0.007}\\
$v_{\textrm{max}}$&     65&    376& -13139.65& -13808.60& -12137.50& -13429.00&   7045.04&   6595.68&     0.108&     0.518&      2.46&      0.303& 0.226\\
$BI$   & 56 & 396 & 1347.91 & 1618.22 & 827.90 & 946.30 & 1447.16 & 1883.32 & 0.112 & 0.549 & 3.05  & 0.397 & 0.355\\ 
$BI_0$ & 72 & 466 & 1431.06 & 1578.10 & 925.30 & 780.40 & 1458.56 & 1909.94 & 0.140 & 0.162 & 11.57 & 0.161 & 0.259\\ 

\hline
   \end{tabular}
\raggedright{
  A comparison of several RL and RQ BAL spectral properties.  $\alpha_{\rm{CIV}}$ and $\alpha_{\rm{MgII}}$ are the UV continuum spectral indices in windows on either side of the \CIV\ and \MgII\ emission lines, respectively.  EW$_{\rm{FeII}}$ is the equivalent width of the iron template used around the \MgII\ emission line.  $v_{\textrm{off}}$ indicates the offset of the peak of the emission lines relative to the systemic velocity, in km s$^{-1}$.  $v_{\textrm{min, G}}$ and $v_{\textrm{max, G}}$ are the minimum and maximum BAL absorption velocities from G09, while $v_{\rm{min}}$ and $v_{\rm{max}}$ are the minimum and maximum BAL absorption velocities from our own visual inspections.  ${BI}$ is the traditional balnicity index, integrated from 2000 km s$^{-1}$, and ${BI_0}$ is the modified balnicity index, integrated from 0 km s$^{-1}$; both are measured for \CIV.  All EW measures are in units of \AA\ (in the rest-frame), and FWHM measures are in km s$^{-1}$ . The mean ($\mu$), median ($Med$), standard deviation ($\sigma$), and number ($N$) of RL BAL or RQ BAL sources with a given measurement are presented in columns 2-9. The final five columns show results of statistical tests comparing the RL and the RQ BAL distributions, using KS tests, Monte-Carlo resampling, $\chi^{2}$ tests, and RS tests.  $P$-values of properties that may be significantly different ($P \le 0.05$) are given in bold. 
  }

\end{table*}

\begin{figure*}
\centering
\hspace{0cm}
   \includegraphics[width=12cm]{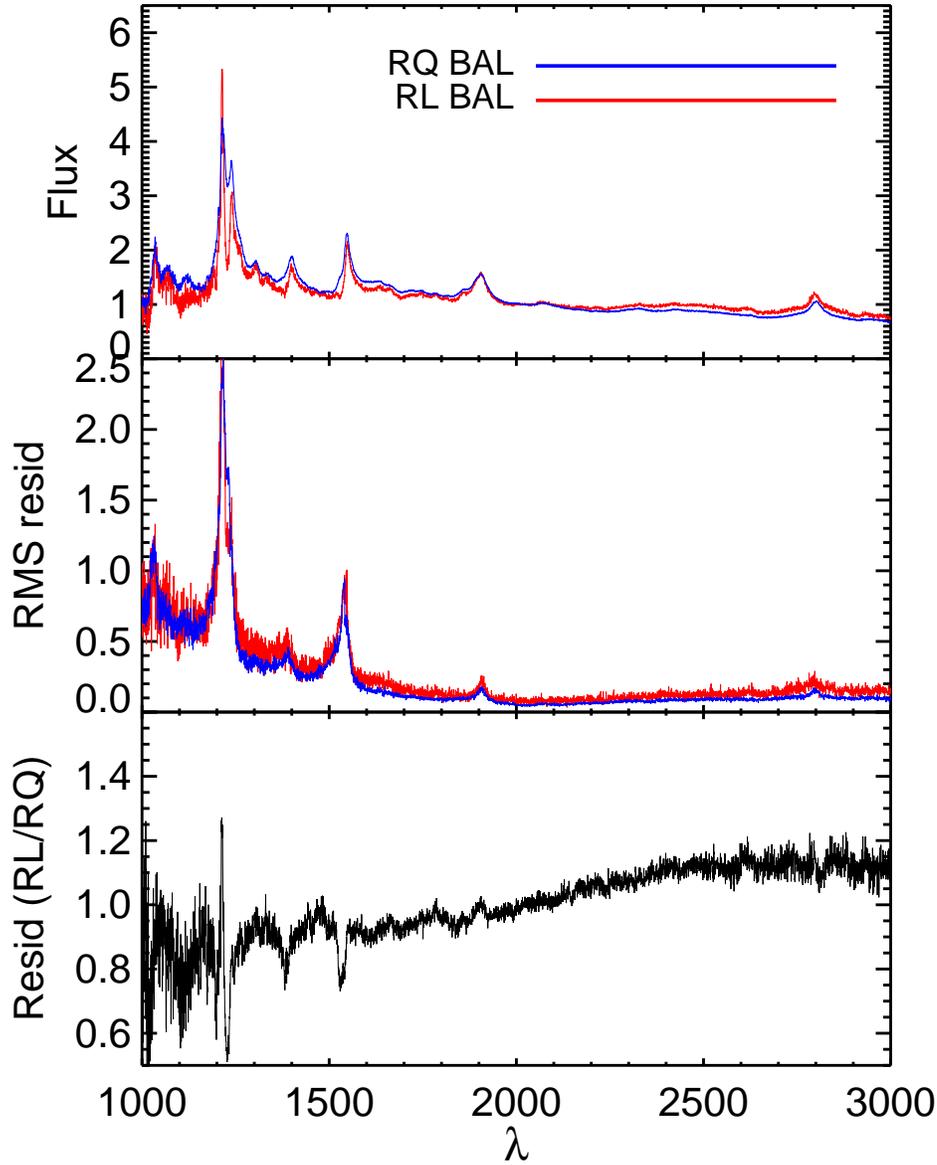}
   \vspace{0 cm}
  \caption{The top panel shows the composite spectra of the RL and RQ BAL samples (after normalizing each spectrum in the wavelength range 2000 to 2050 \AA), the middle panel shows the RMS residual spectra, and the bottom panel shows the RL composite divided by the RQ composite.  While some differences are apparent in the top and bottom panels, the RMS spectra show that the scatter in individual spectra in a given sample is generally comparable in size to these differences.\label{fig:composite}}
\end{figure*}

\section{DISCUSSION}
The key result of this work is the lack of strong differences in the UV spectral properties of RL and RQ BAL quasars, as well as between their fundamental physical properties, confirming the findings from the rest-frame optical of \citet{Bruni14}.  Though Table 2 shows that there are some differences, many of them are interrelated.  We discuss each of them in turn here.

At first glance, it seems that the difference in \FeII\ EW should not be very surprising.  It is known that there is a separation in EV1 between RL and RQ quasars, and the strength of \FeII\ emission is a key contributor to EV1.  However, the situation here is complex, for two reasons.  First, while optical \FeII\ emission is a strong EV1 indicator, this is not always true for \FeII\ emission in the UV \citep[e.g.][]{Wills85, Verner04}.  Second, BAL quasars generally fall on the opposite end of EV1 compared to RL  non-BAL quasars --- BAL quasars tend to have strong \FeII\ and weak \OIII, while the opposite is true for RL quasars.  However, \citet{Runnoe13} find that RL BAL quasars are more ``BAL-like'' than ``RL-like'' in their EV1 properties, in that they typically have strong optical \FeII\ emission and weak \OIII, just like RQ BAL quasars.

While we find that there is a difference in UV \FeII\ strength between RL and RQ BAL quasars, the difference is the opposite of what is found in optical EV1 studies --- RL BAL quasars have stronger UV \FeII\ emission than RQ BAL quasars, on average.  The standard deviation is quite large in both samples, and so it is possible that this difference is purely due to chance and the spread in EV1 amongst all BAL quasars is not large enough to see the same effect as what is seen for RL and RQ non-BAL quasars.  

It appears from the composite spectra that \FeII\ is slightly stronger on average in the RL objects in these samples, though \FeII\ is strong in both samples compared to non-BAL quasars, and the scatter is large.  It is possible that UV \FeII\ is not a satisfactory EV1 discriminator, and the difference measured here is unrelated to EV1.  Because \CIV\ emission lines are difficult to measure in BAL quasars (see below), we opt not to use \CIV\ properties as an independent EV1 proxy.  Higher signal-to-noise spectra, which could allow accurate measurements of the \CIII/\AlIII\ complex (a strong EV1 indicator in the UV), or infra-red spectra (allowing for measurement of the rest-frame optical \FeII\ strength) are needed to make conclusive statements about the EV1 properties of these samples.

Another difference highlighted in Table 2 may be related to the point above --- the difference between the UV spectral index around \MgII.  Disentangling the continuum and \FeII\ emission during spectral fits is a difficult process.  It could simply be that the template fitting of \FeII\ around \MgII\ is imperfect, and the difference in $\alpha$ around \MgII\ is a manifestation of the differences in \FeII\, or vice versa.  Examination of the composite spectra seems to suggest that the difference in $\alpha$ around \MgII\ is in reality quite small, and only appears different due to different \FeII\ strengths.

We note that the \CIV\ emission-line offset velocity distributions of both of our samples resemble the RQ quasar distribution of \citet{Richards11}, despite the finding in that work that BAL quasars and RL non-BAL quasars tend to have different emission-line offset velocities, on average.  Like the above discussion of EV1 properties, it seems that RL BAL quasars are more ``BAL-like'' than ``RL-like'' in their \CIV\ properties as well.  However, measuring \CIV\ emission lines in BAL quasars is difficult, as discussed below.

The next set of significant differences arise from the \CIV\ EW and FWHM.  However, it is likely that these apparent differences stem from the slightly different velocities in the BAL absorption, particularly the minimum velocities.  As expected, our minimum velocities are systematically lower than those of G09, since our visual inspection is more sensitive to absorption higher on the emission line.  Using our $v_{\textrm{min}}$, we find that RL BAL quasars have slightly larger minimum velocities at $\sim 2\sigma$ level.  However, using the measurements of G09, it appears RL BAL quasars have a \textit{lower} minimum velocity at a similar significance.  Examination of the composite spectra seems to indicate that these differences are simply artifacts of systematic errors in the measurement --- the sharp drop in the \CIV\ emission line occurs at very nearly the same place, certainly within the errors.  The composites do seem to indicate that on average the low-velocity absorption is deeper in the RL BAL quasars.  This difference is of a similar magnitude to the RMS in this region, but it could indicate a jet wind-interaction that results in more low velocity gas in RL objects, in line with differences in maximum velocity as described below. The composite also illustrates why the \CIV\ FWHM and EW appear slightly larger in the RQ objects.  It seems clear that the red side of the \CIV\ lines are very similar in shape (this is confirmed by re-normalizing the spectra to the same peak \CIV\ flux), but the blue side differs because of slightly different average absorption properties.

There may be a difference in maximum absorption velocity as well, at least in the primary, deepest absorption troughs.  This is suggested by the fact that there is a significant difference when using $v_{\textrm{max, G}}$, but not the $v_{\textrm{max}}$ measured here, which is more sensitive to smaller, detached BAL troughs at higher velocity.  This can be interpreted in a few ways.  If a line of sight intersecting a BAL outflow is more probable at larger viewing angles (more equatorial), as in \citet{DiPompeo11a}, this result could be due to a bias from relativistic beaming.  If the RL sample is more likely to be viewed from smaller viewing angles compared to the RQ sample, then the line of sight may cut through the outflow at larger angles and higher velocities in the latter objects.  

It is also possible that there is an interaction between the radio jets and the BAL outflow (at least in a subset of objects), that causes both the maximum outflow velocity to drop, and increase the amount of gas at lower velocities as suggested above.  Interactions between radio jets and outflowing material have been suggested in other works \citep[e.g.,][]{Reynolds09, Kunert10, Doi13}.  It may be that there is only an interaction in objects with polar outflows, which could be tested in a sample where the viewing angle to the RL sample is constrained.  However, \citet{DiPompeo12} found no correlation between radio spectral index (a proxy for viewing angle) and BAL velocities, which makes this explanation less likely.  Finally, if RL and RQ states for individual objects are periodic, and the jets and outflows interact, then the timescales on which jets turn on and off and jets effect the outflow properties are important.  In this scenario, the RQ sample is a mix of objects that have been recently turned off and have been quiet for some time, and this could dilute the measured difference in the effect on the outflow.  Unfortunately, our measurements do not allow us to distinguish between these three scenarios.

Finally, though the KS test does not indicate a difference, the $\chi^2$ test suggests that there is a difference in $L_{\textrm{bol}}$.  This difference stems largely from the difference in the monochromatic luminosity at 1350 \AA\, which is the value used to calculate $L_{\textrm{bol}}$ for objects with $z > 1.9$ in S11.  If $L_{3000}$ is considered, this difference disappears.  Looking at the composite spectra, we can see that the RL BAL quasars are slightly redder than RQ quasars, and this likely explains why there is a difference only at bluer wavelengths.  The intrinsic reddening may be slightly different, but the intrinsic luminosities are the same, as they should be by design.

None of the results here allow us to conclusively discriminate between BAL models, and it is unlikely that larger samples from surveys in the near future will change this, at least for similar studies of the UV spectral properties.  The largest limiting factor to this study is the smaller size of the RL BAL sample.  While new data from SDSS-III are increasing the number of known BAL quasars \citep[particularly at high redshift;][]{Paris14}, a radio survey with far greater depth than FIRST, but over at least a similar area, is needed to find significantly larger samples of RL BAL quasars.  Regardless, given the minor differences identified here via a multitude of statistical tests, it seems unlikely that even sample sizes significantly larger will show strong differences that could constrain models.

However, these results are in line with other studies finding that the overall properties of BAL and non-BAL quasars, whether RL or RQ, are remarkably similar with the exception of their radio power or the presence of BAL features.  This is especially true when considering the study of \citet{DiPompeo12}, which compared the UV spectral properties of the RL sample utilized in this work with a similarly well-matched sample of RL non-BAL quasars.  Virtually no strong differences were found there either.  In fact, when analyzed in conjunction with that comparison, it seems that RL and RQ BAL quasars are even more similar (e.g., they both have stronger UV \FeII\ emission, redder continua, etc.).  It appears that at least some BAL outflows are in the polar, not equatorial directions \citep[e.g.][]{Zhou06, Ghosh07, Dipompeo12b}.  BAL outflows are driven by similar mechanisms in all directions \citep{DiPompeo12}, and given the strong similarities at other wavelengths, these results extend to RQ BAL quasars.  However, despite these observed similarities, the presence of BALs is not completely determined by orientation alone.  The nature of BAL quasars is far more complex than simple one-parameter models suggest.

\section{Conclusions}
We have performed a detailed comparison of the rest-frame UV spectral properties of 73 RL BAL and 473 RQ BAL quasars from the SDSS.  These samples are well matched in redshift as well as luminosity. We use several statistical tests, as well as composite spectra, to search for differences between several continuum, emission line, absorption line, and fundamental physical properties.

Generally, the RL and RQ BAL quasars are not significantly different, and we summarize the marginal differences found here:

\begin{enumerate}
\item There does appear to be a slight difference in \FeII\ emission strength, with RL BAL quasars having slightly stronger UV \FeII\ emission.  This difference is the opposite of what is seen in comparisons of optical \FeII\ emission in RL and RQ non-BAL quasars, suggesting that either this difference is simply due to chance and large scatter, or UV \FeII\ emission behaves differently from the optical \FeII\ emission.

\vspace{0.3cm}

\item While \CIV\ EW and FWHM do appear to differ between the samples, it is more likely that these differences arise from the difficulty in measuring the intrinsic \CIV\ emission lines in BAL quasars combined with subtle differences in velocity structure (see iii).

\vspace{0.3cm}

\item We find some differences between the minimum outflow velocities that are difficult to interpret, because the sense of the difference changes depending on how the measurements are made. But visual inspection of the composite spectra indicates that these statistical differences are not real in any case.  However, the composites suggest that there is more low velocity gas in the RL objects, which could be due to a jet-BAL outflow interaction.

\vspace{0.3cm}

\item The differences in maximum velocity indicate that RQ objects have higher maximum outflow velocity.  These are also difficult to interpret, due to potential biases in the average orientation of the samples, but could also indicate an interaction between the radio jets and the BAL outflows.

\end{enumerate}

Unfortunately, the results here cannot provide direct constraints on various BAL quasar models, but they do confirm that in general the physical properties of RL and RQ BAL quasars are at best only marginally different.  This is additional evidence that suggests that recent results concerning, for example, the orientation distribution of RL BAL quasars are extendable to RQ BAL quasars.

\section*{Acknowledgement}
TBR, MAD, and ADM were partially supported by NASA through ADAP award NNX12AE38G and EPSCoR award NNX11AM18A and by the National Science Foundation through grant numbers 1211096 and 1211112.

\bibliography{references.bib}

\label{lastpage}

\end{document}

%% file: commands_mnras.tex
\newcommand{\CIV}{C\,{\sc iv}}
\newcommand{\MgII}{Mg\,{\sc ii}}

\newcommand{\AlIII}{Al\,{\sc iii}}

\newcommand{\CIII}{C\,{\sc iii}]}
\newcommand{\FeII}{Fe\,{\sc ii}}

\newcommand{\NV}{N\,{\sc v}}

\newcommand{\OIII}{[O\,{\sc iii}]}

\newcommand{\SiIV}{Si\,{\sc iv}}

\newcommand\ion[2]{#1$\;${\small\rmfamily\@Roman{#2}}\relax}%

\def\lsim{\lower0.3em\hbox{$\,\buildrel <\over\sim\,$}}
\def\gsim{\lower0.3em\hbox{$\,\buildrel >\over\sim\,$}}

%% file: rl_rq_bals.bbl
\begin{thebibliography}{59}
\expandafter\ifx\csname natexlab\endcsname\relax\def\natexlab#1{#1}\fi

\bibitem[{{Adelman-McCarthy} {et~al}\mbox{.}(2007){Adelman-McCarthy},
  {Ag{\"u}eros}, {Allam}, {Anderson}, {Anderson}, {Annis}, {Bahcall},
  {Bailer-Jones}, {Baldry}, {Barentine}, {Beers}, {Belokurov}, {Berlind},
  {Bernardi}, {Blanton}, {Bochanski}, {Boroski}, {Bramich}, {Brewington},
  {Brinchmann}, {Brinkmann}, {Brunner}, {Budav{\'a}ri}, {Carey}, {Carliles},
  {Carr}, {Castander}, {Connolly}, {Cool}, {Cunha}, {Csabai}, {Dalcanton},
  {Doi}, {Eisenstein}, {Evans}, {Evans}, {Fan}, {Finkbeiner}, {Friedman},
  {Frieman}, {Fukugita}, {Gillespie}, {Gilmore}, {Glazebrook}, {Gray},
  {Grebel}, {Gunn}, {de Haas}, {Hall}, {Harvanek}, {Hawley}, {Hayes},
  {Heckman}, {Hendry}, {Hennessy}, {Hindsley}, {Hirata}, {Hogan}, {Hogg},
  {Holtzman}, {Ichikawa}, {Ichikawa}, {Ivezi{\'c}}, {Jester}, {Johnston},
  {Jorgensen}, {Juri{\'c}}, {Kauffmann}, {Kent}, {Kleinman}, {Knapp},
  {Kniazev}, {Kron}, {Krzesinski}, {Kuropatkin}, {Lamb}, {Lampeitl}, {Lee},
  {Leger}, {Lima}, {Lin}, {Long}, {Loveday}, {Lupton}, {Mandelbaum}, {Margon},
  {Mart{\'{\i}}nez-Delgado}, {Matsubara}, {McGehee}, {McKay}, {Meiksin},
  {Munn}, {Nakajima}, {Nash}, {Neilsen}, {Newberg}, {Nichol},
  {Nieto-Santisteban}, {Nitta}, {Oyaizu}, {Okamura}, {Ostriker}, {Padmanabhan},
  {Park}, {Peoples}, {Pier}, {Pope}, {Pourbaix}, {Quinn}, {Raddick}, {Re
  Fiorentin}, {Richards}, {Richmond}, {Rix}, {Rockosi}, {Schlegel},
  {Schneider}, {Scranton}, {Seljak}, {Sheldon}, {Shimasaku}, {Silvestri},
  {Smith}, {Smol{\v c}i{\'c}}, {Snedden}, {Stebbins}, {Stoughton}, {Strauss},
  {SubbaRao}, {Suto}, {Szalay}, {Szapudi}, {Szkody}, {Tegmark}, {Thakar},
  {Tremonti}, {Tucker}, {Uomoto}, {Vanden Berk}, {Vandenberg}, {Vidrih},
  {Vogeley}, {Voges}, {Vogt}, {Weinberg}, {West}, {White}, {Wilhite}, {Yanny},
  {Yocum}, {York}, {Zehavi}, {Zibetti}, \& {Zucker}}]{SDSSDR5}
{Adelman-McCarthy} J.~K. {et~al.}, 2007, \apjs, 172, 634

\bibitem[{{Allen} {et~al}\mbox{.}(2011){Allen}, {Hewett}, {Maddox}, {Richards},
  \& {Belokurov}}]{Allen11}
{Allen} J.~T., {Hewett} P.~C., {Maddox} N., {Richards} G.~T., {Belokurov} V.,
  2011, \mnras, 410, 860

\bibitem[{{Antonucci}(1993)}]{Antonucci93}
{Antonucci} R., 1993, \araa, 31, 473

\bibitem[{{Becker} {et~al}\mbox{.}(2000){Becker}, {White}, {Gregg},
  {Brotherton}, {Laurent-Muehleisen}, \& {Arav}}]{Becker00}
{Becker} R.~H., {White} R.~L., {Gregg} M.~D., {Brotherton} M.~S.,
  {Laurent-Muehleisen} S.~A., {Arav} N., 2000, \apj, 538, 72

\bibitem[{{Becker} {et~al}\mbox{.}(2001){Becker}, {White}, {Gregg},
  {Laurent-Muehleisen}, {Brotherton}, {Impey}, {Chaffee}, {Richards},
  {Helfand}, {Lacy}, {Courbin}, \& {Proctor}}]{Becker01}
{Becker} R.~H. {et~al.}, 2001, \apjs, 135, 227

\bibitem[{{Becker}, {White} \& {Helfand}(1995){Becker}, {White}, \&
  {Helfand}}]{BeckerFIRST95}
{Becker} R.~H., {White} R.~L., {Helfand} D.~J., 1995, \apj, 450, 559

\bibitem[{{Boroson} \& {Green}(1992)}]{BorosonGreen92}
{Boroson} T.~A., {Green} R.~F., 1992, \apjs, 80, 109

\bibitem[{{Brotherton} {et~al}\mbox{.}(2005){Brotherton}, {Laurent-Muehleisen},
  {Becker}, {Gregg}, {Telis}, {White}, \& {Shang}}]{Brotherton05}
{Brotherton} M.~S., {Laurent-Muehleisen} S.~A., {Becker} R.~H., {Gregg} M.~D.,
  {Telis} G., {White} R.~L., {Shang} Z., 2005, \aj, 130, 2006

\bibitem[{{Brotherton} {et~al}\mbox{.}(2001){Brotherton}, {Tran}, {Becker},
  {Gregg}, {Laurent-Muehleisen}, \& {White}}]{Brotherton01}
{Brotherton} M.~S., {Tran} H.~D., {Becker} R.~H., {Gregg} M.~D.,
  {Laurent-Muehleisen} S.~A., {White} R.~L., 2001, \apj, 546, 775

\bibitem[{{Brotherton} {et~al}\mbox{.}(1998){Brotherton}, {van Breugel},
  {Smith}, {Boyle}, {Shanks}, {Croom}, {Miller}, \& {Becker}}]{Brotherton98}
{Brotherton} M.~S., {van Breugel} W., {Smith} R.~J., {Boyle} B.~J., {Shanks}
  T., {Croom} S.~M., {Miller} L., {Becker} R.~H., 1998, \apjl, 505, L7

\bibitem[{Bruni {et~al}\mbox{.}(2014)Bruni, Gonz{\'a}lez-Serrano, Pedani, Benn,
  Mack, Holt, Montenegro-Montes, \& Jim{\'e}nez-Luj{\'a}n}]{Bruni14}
Bruni G., Gonz{\'a}lez-Serrano J.~I., Pedani M., Benn C.~R., Mack K.-H., Holt
  J., Montenegro-Montes F.~M., Jim{\'e}nez-Luj{\'a}n F., 2014, arXiv, 7987

\bibitem[{{Bruni} {et~al}\mbox{.}(2012){Bruni}, {Mack}, {Salerno},
  {Montenegro-Montes}, {Carballo}, {Benn}, {Gonz{\'a}lez-Serrano}, {Holt}, \&
  {Jim{\'e}nez-Luj{\'a}n}}]{Bruni12}
{Bruni} G. {et~al.}, 2012, \aap, 542, A13

\bibitem[{{Cao Orjales} {et~al}\mbox{.}(2012){Cao Orjales}, {Stevens},
  {Jarvis}, {Smith}, {Hardcastle}, {Auld}, {Baes}, {Cava}, {Clements},
  {Cooray}, {Coppin}, {Dariush}, {De Zotti}, {Dunne}, {Dye}, {Eales},
  {Hopwood}, {Hoyos}, {Ibar}, {Ivison}, {Maddox}, {Page}, \&
  {Valiante}}]{Cao12}
{Cao Orjales} J.~M. {et~al.}, 2012, \mnras, 427, 1209

\bibitem[{{DiPompeo} {et~al}\mbox{.}(2012){DiPompeo}, {Brotherton}, {Cales}, \&
  {Runnoe}}]{DiPompeo12}
{DiPompeo} M.~A., {Brotherton} M.~S., {Cales} S.~L., {Runnoe} J.~C., 2012,
  \mnras, 427, 1135

\bibitem[{{DiPompeo}, {Brotherton} \& {De Breuck}(2011){DiPompeo},
  {Brotherton}, \& {De Breuck}}]{DiPompeo11a}
{DiPompeo} M.~A., {Brotherton} M.~S., {De Breuck} C., 2011, \apjs, 193, 9

\bibitem[{{DiPompeo}, {Brotherton} \& {De Breuck}(2012){DiPompeo},
  {Brotherton}, \& {De Breuck}}]{Dipompeo12b}
{DiPompeo} M.~A., {Brotherton} M.~S., {De Breuck} C., 2012, \apj, 752, 6

\bibitem[{{DiPompeo} {et~al}\mbox{.}(2011){DiPompeo}, {Brotherton}, {De
  Breuck}, \& {Laurent-Muehleisen}}]{DiPompeo11b}
{DiPompeo} M.~A., {Brotherton} M.~S., {De Breuck} C., {Laurent-Muehleisen} S.,
  2011, \apj, 743, 71

\bibitem[{{DiPompeo} {et~al}\mbox{.}(2013){DiPompeo}, {Runnoe}, {Brotherton},
  \& {Myers}}]{DiPompeo13}
{DiPompeo} M.~A., {Runnoe} J.~C., {Brotherton} M.~S., {Myers} A.~D., 2013,
  \apj, 762, 111

\bibitem[{{Doi} {et~al}\mbox{.}(2013){Doi}, {Murata}, {Mochizuki}, {Takeuchi},
  {Asada}, {Hayashi}, {Nagai}, {Shibata}, {Oyama}, {Jike}, {Fujisawa},
  {Sugiyama}, {Ogawa}, {Kimura}, {Honma}, {Kobayashi}, \& {Koyama}}]{Doi13}
{Doi} A. {et~al.}, 2013, PASJ, 65, 57

\bibitem[{{Elvis}(2000)}]{Elvis00}
{Elvis} M., 2000, \apj, 545, 63

\bibitem[{{Filiz Ak} {et~al}\mbox{.}(2012){Filiz Ak}, {Brandt}, {Hall},
  {Schneider}, {Anderson}, {Gibson}, {Lundgren}, {Myers}, {Petitjean}, {Ross},
  {Shen}, {York}, {Bizyaev}, {Brinkmann}, {Malanushenko}, {Oravetz}, {Pan},
  {Simmons}, \& {Weaver}}]{FilizAk12}
{Filiz Ak} N. {et~al.}, 2012, \apj, 757, 114

\bibitem[{{Filiz Ak} {et~al}\mbox{.}(2013){Filiz Ak}, {Brandt}, {Hall},
  {Schneider}, {Anderson}, {Hamann}, {Lundgren}, {Myers}, {P{\^a}ris},
  {Petitjean}, {Ross}, {Shen}, \& {York}}]{FilizAk13}
{Filiz Ak} N. {et~al.}, 2013, \apj, 777, 168

\bibitem[{{Gallagher} {et~al}\mbox{.}(2007){Gallagher}, {Hines}, {Blaylock},
  {Priddey}, {Brandt}, \& {Egami}}]{Gallagher07}
{Gallagher} S.~C., {Hines} D.~C., {Blaylock} M., {Priddey} R.~S., {Brandt}
  W.~N., {Egami} E.~E., 2007, \apj, 665, 157

\bibitem[{{Ghosh} \& {Punsly}(2007)}]{Ghosh07}
{Ghosh} K.~K., {Punsly} B., 2007, \apjl, 661, L139

\bibitem[{{Gibson} {et~al}\mbox{.}(2008){Gibson}, {Brandt}, {Schneider}, \&
  {Gallagher}}]{Gibson08}
{Gibson} R.~R., {Brandt} W.~N., {Schneider} D.~P., {Gallagher} S.~C., 2008,
  \apj, 675, 985

\bibitem[{{Gibson} {et~al}\mbox{.}(2009){Gibson}, {Jiang}, {Brandt}, {Hall},
  {Shen}, {Wu}, {Anderson}, {Schneider}, {Vanden Berk}, {Gallagher}, {Fan}, \&
  {York}}]{GibsonCatalog09}
{Gibson} R.~R. {et~al.}, 2009, \apj, 692, 758

\bibitem[{{Goldschmidt} {et~al}\mbox{.}(1999){Goldschmidt}, {Kukula}, {Miller},
  \& {Dunlop}}]{Goldschmidt99}
{Goldschmidt} P., {Kukula} M.~J., {Miller} L., {Dunlop} J.~S., 1999, \apj, 511,
  612

\bibitem[{{Gregg}, {Becker} \& {de Vries}(2006){Gregg}, {Becker}, \& {de
  Vries}}]{Gregg06}
{Gregg} M.~D., {Becker} R.~H., {de Vries} W., 2006, \apj, 641, 210

\bibitem[{{Hopkins} \& {Elvis}(2010)}]{HopkinsElvis10}
{Hopkins} P.~F., {Elvis} M., 2010, \mnras, 401, 7

\bibitem[{{Hopkins} {et~al}\mbox{.}(2006){Hopkins}, {Hernquist}, {Cox}, {Di
  Matteo}, {Robertson}, \& {Springel}}]{Hopkins06}
{Hopkins} P.~F., {Hernquist} L., {Cox} T.~J., {Di Matteo} T., {Robertson} B.,
  {Springel} V., 2006, \apjs, 163, 1

\bibitem[{{Jiang} {et~al}\mbox{.}(2007){Jiang}, {Fan}, {Ivezi{\'c}},
  {Richards}, {Schneider}, {Strauss}, \& {Kelly}}]{Jiang07}
{Jiang} L., {Fan} X., {Ivezi{\'c}} {\v Z}., {Richards} G.~T., {Schneider}
  D.~P., {Strauss} M.~A., {Kelly} B.~C., 2007, \apj, 656, 680

\bibitem[{{Knigge} {et~al}\mbox{.}(2008){Knigge}, {Scaringi}, {Goad}, \&
  {Cottis}}]{Knigge08}
{Knigge} C., {Scaringi} S., {Goad} M.~R., {Cottis} C.~E., 2008, \mnras, 386,
  1426

\bibitem[{{Komatsu} {et~al}\mbox{.}(2011){Komatsu}, {Smith}, {Dunkley},
  {Bennett}, {Gold}, {Hinshaw}, {Jarosik}, {Larson}, {Nolta}, {Page},
  {Spergel}, {Halpern}, {Hill}, {Kogut}, {Limon}, {Meyer}, {Odegard}, {Tucker},
  {Weiland}, {Wollack}, \& {Wright}}]{Komatsu11}
{Komatsu} E. {et~al.}, 2011, \apjs, 192, 18

\bibitem[{{Kunert-Bajraszewska} {et~al}\mbox{.}(2010){Kunert-Bajraszewska},
  {Janiuk}, {Gawro{\'n}ski}, \& {Siemiginowska}}]{Kunert10}
{Kunert-Bajraszewska} M., {Janiuk} A., {Gawro{\'n}ski} M.~P., {Siemiginowska}
  A., 2010, \apj, 718, 1345

\bibitem[{{Kunert-Bajraszewska} {et~al}\mbox{.}(2009){Kunert-Bajraszewska},
  {Siemiginowska}, {Katarzy{\'n}ski}, \& {Janiuk}}]{Kunert09}
{Kunert-Bajraszewska} M., {Siemiginowska} A., {Katarzy{\'n}ski} K., {Janiuk}
  A., 2009, \apj, 705, 1356

\bibitem[{{Leighly} {et~al}\mbox{.}(2014){Leighly}, {Terndrup}, {Baron},
  {Lucy}, {Dietrich}, \& {Gallagher}}]{Leighly14}
{Leighly} K.~M., {Terndrup} D.~M., {Baron} E., {Lucy} A.~B., {Dietrich} M.,
  {Gallagher} S.~C., 2014, \apj, 788, 123

\bibitem[{{Miller}, {Peacock} \& {Mead}(1990){Miller}, {Peacock}, \&
  {Mead}}]{MillerPeacockMead90}
{Miller} L., {Peacock} J.~A., {Mead} A.~R.~G., 1990, \mnras, 244, 207

\bibitem[{{Montenegro-Montes} {et~al}\mbox{.}(2008){Montenegro-Montes}, {Mack},
  {Vigotti}, {Benn}, {Carballo}, {Gonz{\'a}lez-Serrano}, {Holt}, \&
  {Jim{\'e}nez-Luj{\'a}n}}]{MontenegroMontes08}
{Montenegro-Montes} F.~M., {Mack} K.-H., {Vigotti} M., {Benn} C.~R., {Carballo}
  R., {Gonz{\'a}lez-Serrano} J.~I., {Holt} J., {Jim{\'e}nez-Luj{\'a}n} F.,
  2008, \mnras, 388, 1853

\bibitem[{{P{\^a}ris} {et~al}\mbox{.}(2014){P{\^a}ris}, {Petitjean}, {Aubourg},
  {Ross}, {Myers}, {Streblyanska}, {Bailey}, {Hall}, {Strauss}, {Anderson},
  {Bizyaev}, {Borde}, {Brinkmann}, {Bovy}, {Brandt}, {Brewington},
  {Brownstein}, {Cook}, {Ebelke}, {Fan}, {Filiz Ak}, {Finley}, {Font-Ribera},
  {Ge}, {Hamann}, {Ho}, {Jiang}, {Kinemuchi}, {Malanushenko}, {Malanushenko},
  {Marchante}, {McGreer}, {McMahon}, {Miralda-Escud{\'e}}, {Muna},
  {Noterdaeme}, {Oravetz}, {Palanque-Delabrouille}, {Pan}, {Perez-Fournon},
  {Pieri}, {Riffel}, {Schlegel}, {Schneider}, {Simmons}, {Viel}, {Weaver},
  {Wood-Vasey}, {Y{\`e}che}, \& {York}}]{Paris14}
{P{\^a}ris} I. {et~al.}, 2014, \aap, 563, A54

\bibitem[{Peterson(2003)}]{Peterson03}
Peterson B.~M., 2003, An Introduction to Active Galactic Nuclei. Cambridge,
  Univ. Press, Cambridge

\bibitem[{{Reynolds} {et~al}\mbox{.}(2009){Reynolds}, {Punsly}, {Kharb},
  {O'Dea}, \& {Wrobel}}]{Reynolds09}
{Reynolds} C., {Punsly} B., {Kharb} P., {O'Dea} C.~P., {Wrobel} J., 2009, \apj,
  706, 851

\bibitem[{{Richards} {et~al}\mbox{.}(2011){Richards}, {Kruczek}, {Gallagher},
  {Hall}, {Hewett}, {Leighly}, {Deo}, {Kratzer}, \& {Shen}}]{Richards11}
{Richards} G.~T. {et~al.}, 2011, \aj, 141, 167

\bibitem[{{Runnoe} {et~al}\mbox{.}(2013){Runnoe}, {Ganguly}, {Brotherton}, \&
  {DiPompeo}}]{Runnoe13}
{Runnoe} J.~C., {Ganguly} R., {Brotherton} M.~S., {DiPompeo} M.~A., 2013,
  \mnras, 433, 1778

\bibitem[{{Scannapieco} \& {Oh}(2004)}]{Scannapieco04}
{Scannapieco} E., {Oh} S.~P., 2004, \apj, 608, 62

\bibitem[{{Shankar}, {Dai} \& {Sivakoff}(2008){Shankar}, {Dai}, \&
  {Sivakoff}}]{Shankar08}
{Shankar} F., {Dai} X., {Sivakoff} G.~R., 2008, \apj, 687, 859

\bibitem[{{Shen} {et~al}\mbox{.}(2011){Shen}, {Richards}, {Strauss}, {Hall},
  {Schneider}, {Snedden}, {Bizyaev}, {Brewington}, {Malanushenko},
  {Malanushenko}, {Oravetz}, {Pan}, \& {Simmons}}]{ShenCatalog11}
{Shen} Y. {et~al.}, 2011, \apjs, 194, 45

\bibitem[{{Singal} {et~al}\mbox{.}(2013){Singal}, {Petrosian}, {Stawarz}, \&
  {Lawrence}}]{Singal13}
{Singal} J., {Petrosian} V., {Stawarz} {\L}., {Lawrence} A., 2013, \apj, 764,
  43

\bibitem[{{Sprayberry} \& {Foltz}(1992)}]{Sprayberry92}
{Sprayberry} D., {Foltz} C.~B., 1992, \apj, 390, 39

\bibitem[{{Stocke} {et~al}\mbox{.}(1992){Stocke}, {Morris}, {Weymann}, \&
  {Foltz}}]{Stocke92}
{Stocke} J.~T., {Morris} S.~L., {Weymann} R.~J., {Foltz} C.~B., 1992, \apj,
  396, 487

\bibitem[{{Urry} \& {Padovani}(1995)}]{UrryPadovani95}
{Urry} C.~M., {Padovani} P., 1995, \pasp, 107, 803

\bibitem[{{Vernaleo} \& {Reynolds}(2006)}]{Vernaleo06}
{Vernaleo} J.~C., {Reynolds} C.~S., 2006, \apj, 645, 83

\bibitem[{{Verner} {et~al}\mbox{.}(2004){Verner}, {Bruhweiler}, {Verner},
  {Johansson}, {Kallman}, \& {Gull}}]{Verner04}
{Verner} E., {Bruhweiler} F., {Verner} D., {Johansson} S., {Kallman} T., {Gull}
  T., 2004, \apj, 611, 780

\bibitem[{{Vestergaard} \& {Wilkes}(2001)}]{Vestergaard01}
{Vestergaard} M., {Wilkes} B.~J., 2001, \apjs, 134, 1

\bibitem[{{Welling} {et~al}\mbox{.}(2014){Welling}, {Miller}, {Brandt},
  {Capellupo}, \& {Gibson}}]{Welling14}
{Welling} C.~A., {Miller} B.~P., {Brandt} W.~N., {Capellupo} D.~M., {Gibson}
  R.~R., 2014, \mnras, 440, 2474

\bibitem[{{Weymann} {et~al}\mbox{.}(1991){Weymann}, {Morris}, {Foltz}, \&
  {Hewett}}]{Weymann91}
{Weymann} R.~J., {Morris} S.~L., {Foltz} C.~B., {Hewett} P.~C., 1991, \apj,
  373, 23

\bibitem[{{Willott}, {Rawlings} \& {Grimes}(2003){Willott}, {Rawlings}, \&
  {Grimes}}]{Willott03}
{Willott} C.~J., {Rawlings} S., {Grimes} J.~A., 2003, \apj, 598, 909

\bibitem[{{Wills} \& {Brotherton}(1995)}]{Wills95}
{Wills} B.~J., {Brotherton} M.~S., 1995, \apjl, 448, L81

\bibitem[{{Wills}, {Netzer} \& {Wills}(1985){Wills}, {Netzer}, \&
  {Wills}}]{Wills85}
{Wills} B.~J., {Netzer} H., {Wills} D., 1985, \apj, 288, 94

\bibitem[{{Zhou} {et~al}\mbox{.}(2006){Zhou}, {Wang}, {Wang}, {Wang}, {Yuan},
  \& {Lu}}]{Zhou06}
{Zhou} H., {Wang} T., {Wang} H., {Wang} J., {Yuan} W., {Lu} Y., 2006, \apj,
  639, 716

\end{thebibliography}
